\documentclass[iop,apjl]{emulateapj}
\usepackage{natbib}
\begin{document}

\newcommand{\kms}{\ensuremath{\mathrm{km}\,\mathrm{s}^{-1}}}
\newcommand{\etal}{et al.}
\newcommand{\LCDM}{$\Lambda$CDM}

%\shorttitile{Correlated Residuals}
%\shortauthors{McGaugh \& Wolf}

\title{Local Group Dwarf Spheroidals: Correlated Deviations from the Baryonic Tully-Fisher Relation}

%\author{Stacy S. McGaugh\altaffilmark{1}, Joe Wolf\altaffilmark{2}}

%\altaffiltext{1}{Department of Astronomy, University of Maryland, College Park, MD 20742-2421, USA; 
%ssm@astro.umd.edu}
%\altaffiltext{2}{Center for Cosmology, Department of Physics and Astronomy, University of California, 
%  Irvine, CA 92697-4575}

\author{Stacy S. McGaugh} 

\affil{Department of Astronomy, University of Maryland, College Park, MD 20742-2421, USA; 
ssm@astro.umd.edu}

\and

\author{Joe Wolf}

\affil{Center for Cosmology, Department of Physics and Astronomy, University of California, 
  Irvine, CA 92697-4575, USA; wolfj@uci.edu}

\begin{abstract}
Local Group dwarf spheroidal satellite galaxies are the faintest extragalactic stellar systems known.
We examine recent data for these objects in the plane of the Baryonic Tully-Fisher Relation 
(BTFR).  While some dwarf spheroidals adhere to the BTFR, others deviate substantially.  
We examine the residuals from the BTFR and find that they are not random.
The residuals correlate with luminosity, size, metallicity, ellipticity, and susceptibility of the 
dwarfs to tidal disruption in the sense that
fainter, more elliptical, and tidally more susceptible dwarfs deviate farther from the BTFR.
These correlations disfavor stochastic processes and suggest a role for tidal effects.
We identify a test to distinguish between \LCDM\ and MOND based
on the orbits of the dwarf satellites of the Milky Way and how stars are lost from them. 
\end{abstract}


\section{Introduction}

Recent years have seen enormous progress in the discovery and measurement of the tiny
dwarf satellite galaxies of the Local Group.  These include the long known, ``classical''
dwarf spheroidal galaxies \citep{mateo} as well as the more recently discovered ``ultrafaint''
dwarfs and the satellites of M31 \citep{willman,zucker,grill06,maj07,belokurov,grill09,martin}.  
In addition to identifying these systems, kinematic data
from measuring the velocities of individual stars has become available for many systems.
These now consist of thousands of individual stars for the classical dwarfs \citep{walker07},
with rapidly improving data for the other types of systems \citep{SG,jason}.  

The Local Group dwarfs appear to be the most dark matter dominated objects in the universe
\citep{mateo,wilk02,gerry,SG,koch,strigari,walker,boom}, consistent with the trend of dark matter domination
increasing with decreasing surface brightness \citep{dBMH96,MdB98a}.
As such, they provide a unique probe of structure formation at the smallest accessible scales.
They presumably reside in the sub-halos thought to inhabit the large dark matter halos of 
galaxies like the Milky Way (MW) and M31.  If so, physical processes specific to their environment, 
like tidal disruption and ram pressure stripping,
may play a role in the evolution of the luminous content of
the dwarfs as they orbit the primary structure.  It is therefore interesting to investigate whether,
and the extent to which, the dwarfs obey the scaling relations established for brighter galaxies.

In this paper we investigate how the dwarfs behave in the Baryonic Tully-Fisher plane.  
Rotating disk galaxies define a tight relation between baryonic mass and rotation velocity
\citep{btforig,verhTF}.  This has recently been extended \citep{stark,trach} to low rotation velocities 
($\sim 20\;\mathrm{km}\,\mathrm{s}^{-1}$) comparable to the Local Group dwarf spheroidals.
An obvious question is whether the Local Group dwarfs that are satellites
of the MW and M31 continue this relation.

\section{Data}

We adopt the data compilation of \citet{boom} for the MW satellites augmented by the recent
data for M31 dwarfs presented by \citet{jason}.  These are small, pressure supported systems
\citep{mateo,norot}
whose baryonic mass is in the form of stars.\footnote{Leo T is the one exception to this,
with a non-negligible amount of HI gas \citep{LeoTgas}.  We include the gas mass in the baryonic sum.}
In order to place them
on the same Baryonic Tully-Fisher Relation (BTFR) as rotating disks we need to 
estimate stellar mass and rotation velocity in a consistent manner.

\begin{deluxetable*}{lcccccccr}
\tabletypesize{\scriptsize}
%\rotate
\tablewidth{0pt}
\tablecaption{Baryonic Tully-Fisher Data and Residuals}
\tablehead{
\colhead{Dwarf} & \colhead{$D$} & \colhead{$V_c$} &
\colhead{$\Upsilon_*^V$} & \colhead{$\log M_b$} &
\colhead{$\log F_b$} & \colhead{[Fe/H]} & \colhead{$\delta$[Fe/H]} & \colhead{Refs.} 
%\\ & \multicolumn{2}{c}{(kpc)} & \multicolumn{2}{c}{(km s$^{-1}$)} & \colhead{($M_{\sun}/L_{\sun}$)} &
% \colhead{($\log M_{\sun}$)} & & & & & &
}
\startdata
Carina        &	$101   \pm\phn5$&	$11.1       \pm0.4$&	1.0 &	$5.64\pm0.41$&	$-$0.18 &	$-$1.80   & 0.30    & 1,3,10,12 \\
Draco         &	$\phn76\pm\phn5$&	$17.5       \pm0.9$&	1.5 &	$5.50\pm0.42$&	$-$1.12 &	$-$1.99   & 0.32    & 1,4,10,12 \\
Fornax        &	$138   \pm\phn8$&	$18.5       \pm0.3$&	1.1 &	$7.27\pm0.42$&	\phs0.55 &	$-$1.29   & 0.46    & 1,3,10,12 \\
Leo I         &	$250   \pm30   $&	$15.5       \pm0.7$&	0.9 &	$6.65\pm0.42$&	\phs0.24 &	$-$1.31   & 0.25    & 1,5,10,12 \\
Leo II        &	$205   \pm12   $&	$11.4       \pm0.8$&	1.4 &	$6.03\pm0.42$&	\phs0.16 &	$-$1.74   & 0.23    & 1,6,10,12 \\
Sculptor      &	$\phn79\pm\phn4$&	$15.5       \pm0.3$&	1.4 &	$6.56\pm0.42$&	\phs0.14 &	$-$1.81   & 0.34    & 1,3,10,12 \\
Sextans       &	$\phn86\pm\phn4$&	$12.3       \pm0.5$&	1.3 &	$5.89\pm0.42$&	$-$0.13 &	$-$2.07   & 0.36    & 1,3,10,12 \\
Ursa Minor    &	$\phn66\pm\phn3$&	$20.0       \pm1.1$&	1.5 &	$5.76\pm0.43$&	$-$1.09 &	$-$2.03   & 0.32    & 1,4,10,12 \\
Bo\"otes I    &	$\phn66\pm\phn3$&	$15.7       \pm3.9$&	1.1 &	$4.49\pm0.41$&	$-$1.94 &	$-$2.50   & \nodata & 1,7,11,12 \\
Canes Ven.~I  &	$218   \pm10   $&	$13.2       \pm0.8$&	1.3 &	$5.48\pm0.41$&	$-$0.66 &	$-$2.08   & 0.46    & 1,8,11,12 \\
Canes Ven.~II &	$160   \pm\phn4$&	$\phn7.9    \pm1.8$&	1.0 &	$3.90\pm0.45$&	$-$1.35 &	$-$2.19   & 0.58    & 1,8,11,12 \\
Coma Ber.     &	$\phn45\pm\phn4$&	$\phn8.0    \pm1.4$&	1.3 &	$3.68\pm0.45$&	$-$1.58 &	$-$2.53   & 0.45    & 1,8,11,12 \\
Hercules      &	$132   \pm12   $&	$\phn8.9    \pm1.6$&	1.0 &	$4.05\pm0.43$&	$-$1.39 &	$-$2.58   & 0.51    & 1,8,11,12 \\
Leo IV        &	$160   \pm15   $&	$\phn5.8    \pm3.0$&	1.0 &	$3.94\pm0.46$&	$-$0.75 &	$-$2.58   & 0.75    & 1,8,11,12 \\
Leo T         &	$407   \pm38   $&	$13.5       \pm2.8$&	1.3 &	$5.92\pm0.40$&	$-$0.25 &	$-$2.02   & 0.54    & 1,8,12 \\
Segue 1       &	$\phn28\pm\phn2$&	$\phn7.4    \pm2.0$&	1.8 &	$2.78\pm0.50$&	$-$2.35 &	$-$3.30   & 0.50    & 1,9,11,12 \\
Ursa Major I  &	$\phn97\pm\phn4$&	$13.1       \pm1.8$&	1.4 &	$4.29\pm0.42$&	$-$1.83 &	$-$2.29   & 0.54    & 1,8,11,12 \\
Ursa Major II &	$\phn36\pm\phn5$&	$11.7       \pm2.5$&	1.4 &	$3.75\pm0.45$&	$-$2.17 &	$-$2.44   & 0.57    & 1,8,11,12 \\
Willman 1     &	$\phn43\pm\phn7$&	$\phn6.9    \pm1.6$&	1.5 &	$3.18\pm0.50$&	$-$1.84 &	\nodata & \nodata & 1,11,12 \\
And I         &	$\phn57\pm24   $&	$18.4       \pm1.9$&	1.3 &	$6.77\pm0.40$&	\phs0.06 &	$-$1.45   & 0.37    & 2,12 \\
And II        &	$194   \pm18   $&	$12.6       \pm1.4$&	1.3 &	$7.09\pm0.41$&	\phs1.03 &	$-$1.64   & 0.34    & 2,12 \\
And III       &	$\phn75\pm24   $&	$\phn8.1    \pm3.1$&	1.3 &	$6.11\pm0.42$&	\phs0.82 &	$-$1.78   & 0.27    & 2,12 \\
And VII       &	$215   \pm35   $&	$16.8       \pm2.8$&	1.3 &	$7.37\pm0.42$&	\phs0.82 &	$-$1.40   & 0.30    & 2,12 \\
And X         &	$109   \pm37   $&	$\phn6.8    \pm2.1$&	1.3 &	$5.29\pm0.45$&	\phs0.32 &	$-$1.93   & 0.11    & 2,12 \\
And XIV       &	$186   \pm87   $&	$\phn9.4    \pm2.3$&	1.3 &	$5.37\pm0.45$&	$-$0.16 &	$-$2.26   & 0.31    & 2,12
\enddata
\tablecomments{Columns: (1) Name of the dwarf satellite. 
(2) The distance of the dwarf from the center of its host in kpc.
(3) The circular velocity (equation~\ref{beta}) in km s$^{-1}$.
(4) The $V$-band stellar mass-to-light ratio (in $\mathrm{M}_{\sun}/\mathrm{L}_{\sun}$) assumed to compute baryonic mass.
(5) The logarithm of the baryonic mass ($M_b = \Upsilon_*^V L_V$) in $\mathrm{M}_{\sun}$.
(6) The logarithm of the deviation from the BTFR (equation~\ref{BTFR}): $F_b = M_b/(45 V_c^4)$. 
(7) The mean metallicity and (8) the width of the metallicity distribution.  (9) The source of the data.}
\tablerefs{Distances and kinematic data: 1.~\citet{boom} 2.~\citet{jason}.
Metallicity data: 3.~\citet{helmi06Z} 4.~\citet{winnick} 5.~\citet{kockLeoIZ} 6.~\citet{kockLeo2Z}
7.~\citet{Zmartin} 8.~\citet{kirby} 9.~\citet{geha09}. 
Stellar mass-to-light ratios: 10.~\citet{mario} 11.~\citet{martinmaxL}.
Baryonic mass and BTFR residuals: 12.~this work.}
\label{residtab}
\end{deluxetable*}

The stellar populations of the Local Group dwarfs are not simply ancient
single burst populations.  While some are old, others show clear signs of continuing
intermittent star forming episodes \citep{grebel,HGV}.  \citet{mario} have estimated the 
mass-to-light ratios $\Upsilon_*^V$ of the stars of the classical dwarfs from
resolved observations of their stellar content.  \citet{martinmaxL} have done the same
for many of the ultrafaint dwarfs.  We adopt their stellar mass-to-light ratios for the objects
studied.  The mean of each of these samples is 1.3 M$_{\sun}$/L$_{\sun}$ if we adopt the Kroupa IMF
in the tabulation of  \citet{martinmaxL}.  For objects not specifically addressed by these studies, 
namely Leo T and the satellites of M31, 
we use the mean mass-to-light ratio.  These stellar mass-to-light ratios are subject to 
considerable uncertainty, particularly in the IMF.  We therefore adopt a conservative
uncertainty in this quantity of 0.4 dex.  We also propagate the uncertainties in luminosity and
distance, but in most cases the uncertainty in the mass is dominated by that in the 
mass-to-light ratio: $M_b = \Upsilon_*^V L_V$.

\begin{deluxetable*}{lccccccr}
\tabletypesize{\scriptsize}
%\rotate
\tablewidth{0pt}
\tablecaption{Dwarf Tidal Radii and Ellipticities}
\tablehead{
\colhead{Dwarf} & \colhead{$r_{1/2}$} & \colhead{$r_{t,phot}$} & \colhead{$r_{t,D}$} &
\colhead{$r_{t,M}$} & \colhead{$\gamma$} & \colhead{$\epsilon$} & \colhead{Refs.}
}
\startdata
Carina            &	$   \phn334\pm\phn37$ &	$\phn846 \pm \phn106$              &	$\phn4220 \pm \phn380$ &	$      \phn1230\pm\phn250$    &	$\phn10\phd\phn\pm \phn5\phd\phn$ &	$0.33\pm 0.05$         & 1,6,11 \\
Draco             &	$   \phn291\pm\phn14$ &	$\phn997 \pm \phn\phn13$           &	$\phn4520 \pm \phn360$ &	$   \phn\phn840\pm\phn150$    &	$\phn\phn6.9   \pm \phn2.9$ &	$0.31\pm 0.02$         & 1,7,11 \\
Fornax            &	$   \phn944\pm\phn53$ &	$2854 \pm \phn161$                 &	$10350    \pm \phn720$ &	$      \phn5890\pm1160$       &	$\phn22\phd\phn\pm 10\phd\phn$ &	$0.31\pm 0.03$         & 1,6,11 \\
Leo I             &	$   \phn388\pm\phn64$ &	$\phn850 \pm \phn\phn63$           &	$10230    \pm    1680$ &	$      \phn6660\pm1710$       &	$100\phd\phn   \pm 62\phd\phn$ &	$0.21\pm 0.03$         & 1,6,11 \\
Leo II            &	$   \phn233\pm\phn17$ &	$\phn556 \pm \phn\phn28$           &	$\phn6170 \pm \phn540$ &	$      \phn3400\pm\phn610$    &	$\phn78\phd\phn\pm 33\phd\phn$ &	$0.13\pm 0.05$         & 1,6,11 \\
Sculptor          &	$   \phn375\pm\phn54$ &	$1758 \pm \phn115$                 &	$\phn4660 \pm \phn510$ &	$      \phn1940\pm\phn360$    &	$\phn17\phd\phn\pm \phn8\phd\phn$ &	$0.32\pm 0.03$         & 1,6,11 \\
Sextans           &	$      1019\pm\phn62$ &	$4000 \pm 1250   $                 &	$\phn5900 \pm \phn390$ &	$      \phn1270\pm\phn230$    &	$\phn\phn2.0   \pm \phn0.9$ &	$0.35\pm 0.05$         & 1,6,11 \\
Ursa Minor        &	$   \phn588\pm\phn58$ &	$1495 \pm \phn171$                 &	$\phn5640 \pm \phn490$ &	$   \phn\phn890\pm\phn180$    &	$\phn\phn2.6   \pm \phn1.3$ &	$0.56\pm 0.05$         & 1,6,11 \\
Bo\"otes I        &	$   \phn321\pm\phn28$ &	$\phn860\phm{\pm}\phn\phn\phn\phn$ &	$\phn3920 \pm \phn710$ &	$   \phn\phn330\pm\phn\phn60$ &	$\phn\phn1.5   \pm \phn0.6$ &	$0.39\pm 0.06$         & 1,7,11 \\
Canes Ven.~I  &	$   \phn736\pm\phn47$ &	$3000\phm{\pm}\phn\phn\phn\phn$    &	$10400    \pm \phn770$ &	$      \phn2350\pm\phn370$    &	$\phn\phn8.0   \pm \phn3.0$ &	$0.39\pm 0.03$         & 1,7,11 \\
Canes Ven.~II &	$\phn\phn97\pm\phn16$ &	$\phn500\phm{\pm}\phn\phn\phn\phn$ &	$\phn3060 \pm \phn570$ &	$   \phn\phn510\pm\phn150$    &	$\phn17\phd\phn\pm 11\phd\phn$ &	$0.52~^{~+0.10}_{~-0.11}$ & 1,7,11 \\
Coma Ber.    &	$   \phn100\pm\phn13$ &	$\phn240\phm{\pm}\phn\phn\phn\phn$ &	$\phn1300 \pm \phn220$ &	$   \phn\phn120\pm\phn\phn30$ &	$\phn\phn1.9   \pm \phn1.1$ &	$0.36\pm 0.04$ & 1,8,11 \\
Hercules          &	$   \phn304\pm\phn26$ &	$1500\phm{\pm}\phn\phn\phn\phn$    &	$\phn4240 \pm \phn670$ &	$   \phn\phn480\pm\phn120$    &	$\phn\phn2.8   \pm \phn1.6$ &	$0.67\pm 0.03$ & 1,9,11 \\
Leo IV            &	$   \phn151\pm\phn39$ &	$\phn700\phm{\pm}\phn\phn\phn\phn$ &	$\phn2870 \pm    1140$ &	$   \phn\phn530\pm\phn150$    &	$\phn\phn9\phd\phn\pm \phn7\phd\phn$ &	$\lesssim 0.15$ & 1,10,11 \\
Leo T             &	$   \phn152\pm\phn21$ &	$\phn568 \pm \phn118$              &	$\phn9470 \pm    1810$ &	$      \phn6180\pm1000$       &	$137\phd\phn   \pm 23\phd\phn$ &	$0.29~^{~+0.12}_{~-0.14}$ & 1,7,11 \\
Segue 1           &	$\phn\phn38\pm\phn\phn9$ &	$\phn160\phm{\pm}\phn\phn\phn\phn$ &	$\phn\phn630\pm\phn150$ &	$\phn\phn\phn38\pm\phn\phn12$ &	$\phn\phn1.4   \pm \phn0.7$ &	$0.48~^{~+0.10}_{~-0.13}$ & 1,7,11 \\
Ursa Major I      &	$   \phn415\pm\phn60$ &	$1400\phm{\pm}\phn\phn\phn\phn$    &	$\phn4940 \pm \phn690$ &	$   \phn\phn420\pm\phn\phn70$ &	$\phn\phn1.4   \pm \phn0.7$ &	$0.80\pm 0.04$         & 1,7,11 \\
Ursa Major II     &	$   \phn183\pm\phn33$ &	$\phn500\phm{\pm}\phn\phn\phn\phn$ &	$\phn1730 \pm \phn400$ &	$\phn\phn103\pm\phn\phn28$ &	$\phn\phn0.6   \pm \phn0.4$ &	$0.63~^{~+0.03}_{~-0.05}$ & 1,7,11 \\
Willman 1         &	$\phn\phn33\pm\phn\phn8$ &	$\phn110\phm{\pm}\phn\phn\phn\phn$ &	$\phn\phn780\pm\phn210$ &	$\phn\phn\phn79\pm\phn\phn28$ &	$\phn\phn5.3   \pm \phn4.6$ &	$0.47\pm 0.08$         & 1,7,11 \\
And I             &	$   \phn900\pm\phn75$ &	$2530 \pm \phn230$                 &	$25500    \pm    2400$ &	$         17200\pm1600$       &	$118\phd\phn   \pm 40\phd\phn$ &	$0.22\pm 0.04$         & 2,3,11 \\
And II            &	$      1659\pm\phn53$ &	$4170 \pm \phn200$                 &	$22300    \pm    1800$ &	$         19200\pm1900$       &	$\phn56\phd\phn\pm 20\phd\phn$ &	$0.20\pm 0.08$         & 2,3,11 \\
And III           &	$   \phn638\pm\phn77$ &	$2120 \pm \phn360$                 &	$13300    \pm    3600$ &	$         10500\pm1200$       &	$\phn94\phd\phn\pm 43\phd\phn$ &	$0.52\pm 0.02$         & 2,3,11 \\
And VII           &	$      1050\pm\phn60$ &	$4600 \pm \phn430$                 &	$25700    \pm    3200$ &	$         27900\pm3300$       &	$193\phd\phn   \pm 83\phd\phn$ &	$0.13\pm 0.04$         & 2,3,11 \\
And X             &	$   \phn448\pm\phn\phn8$ &	$1480 \pm \phn150$                 &	$10000    \pm    2100$ &	$      \phn5200\pm\phn800$    &	$\phn56\phd\phn\pm 32\phd\phn$ &	\nodata & 2,4,11 \\
And XIV           &	$   \phn461\pm155$ &	$1010 \pm \phn200$                 &	$14500    \pm    4200$ &	$      \phn6900\pm1300$       &	$\phn81\phd\phn\pm 64\phd\phn$ &	$0.31\pm 0.09$  & 2,5,11 
\enddata
\tablecomments{All radii are in parsecs.  
Columns: (1) Name of the dwarf satellite. 
(2) The deprojected 3D half light radius. 
(3) The photometric tidal radius.  Cases lacking error bars are approximate estimates only.
(4) The tidal radius with dark matter from equation~(\ref{RTconventional}) with $m = M_{1/2}$.
(5) The tidal radius in MOND from equation~(\ref{RTMOND}) with $m = M_b$.
(6) The number of orbits a star at the deprojected half light radius completes
for every orbit of the dwarf about the host (equation~\ref{mondadiabat}).
(7) The ellipticity ($\epsilon = 1-b/a$) of the dwarf as projected on the sky.
(8) The source of the data. }
\tablerefs{Deprojected half light and photometric tidal radii: 1.~\citet{boom} 2.~\citet{jason} 3.~\citet{MI06}
4.~\citet{zucker07} 5.~\citet{maj07}.
Ellipticity data: 6.~\citet{mateo} 7.~\citet{martinmaxL}
8.~\citet{munozUMII} 9.~\citet{sandherc} 10.~\citet{sandleo4}.
Tidal radii with dark matter and MOND and $\gamma$: 11.~this work.
}
\label{radtab}
\end{deluxetable*}

The stellar mass is woefully inadequate to explain the observed
velocity dispersions of these systems:  they appear to be dark matter dominated.  We adopt the
line of sight velocity dispersions $\sigma_{los}$ tabulated by \citet{boom} for the Milky Way
satellites and \citet{jason} for the faint satellites of M31.
In order to relate this value to the characteristic circular velocities
of rotating disk galaxies, we assume
\begin{equation}
V_c = \sqrt{3} \sigma_{los}.
\label{beta}
\end{equation}
We take this to be the circular velocity characteristic of the dark matter halo of each dwarf.
This is not precisely the same quantity as measured in disks.  Geometry can cause differences
up to $\sim 20\%$ \citep{MdB98b}, in the sense that flattened systems of the same mass rotate
faster.  In addition,
the radius at which $\sigma_{los}$ is measured is typically modest compared to that of spirals.
Nevertheless, adopting the velocity dispersion measured at the 3D half light radius 
minimizes the uncertainty due to orbital anisotropy \citep{boom} and
provides a lower limit on the characteristic circular velocity at larger radii.
Moreover, the velocity dispersion profiles of the well-observed classical dwarfs tend to be
rather flat \citep{walker07}, so this is the best available estimator of the gravitational potential.
The circular velocities calculated in this manner are listed in Table \ref{residtab}.  We also tabulate
other useful quantities, computed here or collected from the literature, in  Tables \ref{residtab}
and  \ref{radtab}.

\section{Interpretations}

\begin{figure*}
\epsscale{1.0}
\plotone{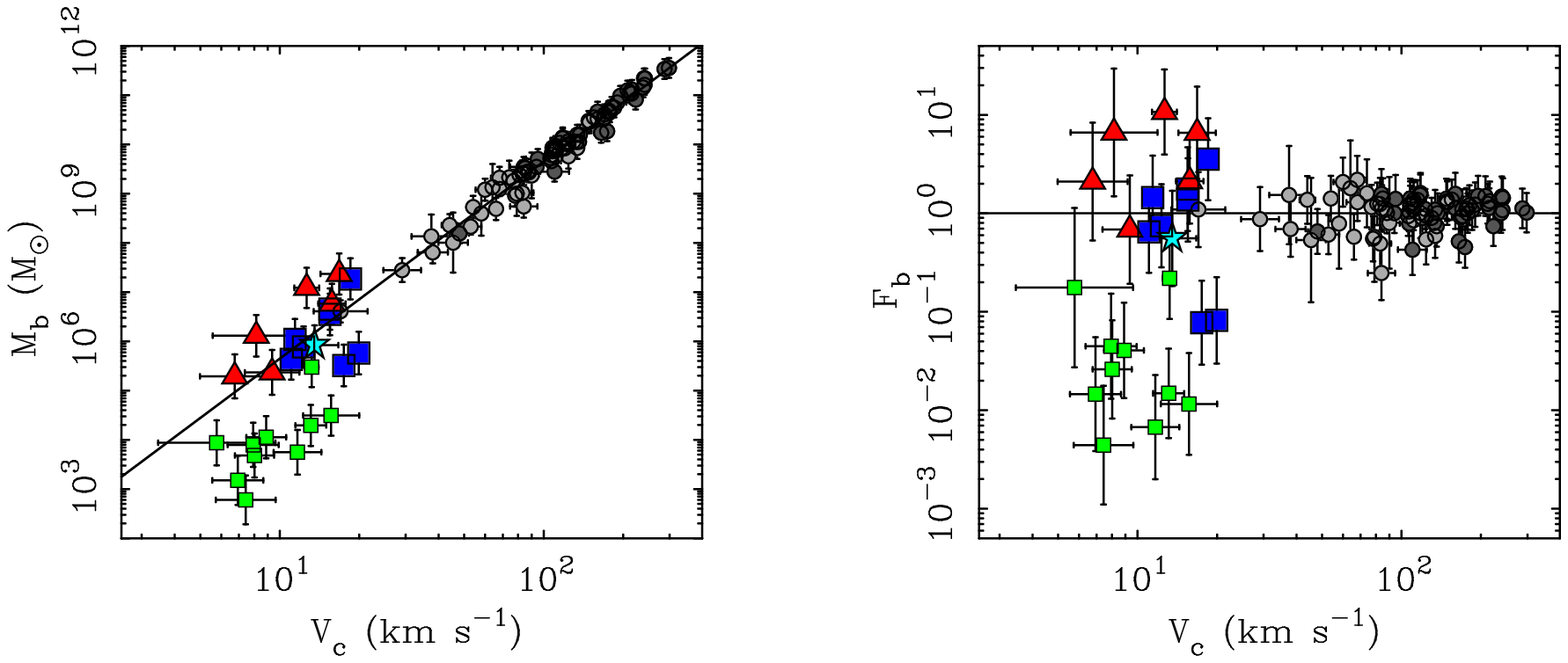}
\caption{The masses and circular velocities of galaxies: the Baryonic Tully-Fisher Relation (left panel) 
and residuals from the BTFR (right panel) obtained by dividing out equation~(\ref{BTFR}) (lines).  
Gray circles are rotating disk galaxies.
Dark gray points are star dominated spirals \citep{M05} while light gray points are gas dominated
disks \citep{stark,trach}.  The circular velocities of non-rotating galaxies are estimated as
$V_c = \sqrt{3} \sigma_{los}$ (equation \ref{beta}).
Large blue squares are the
classical dwarf Spheroidals and the small green squares are the ultrafaint dwarfs as tabulated by
\citet{boom}.  Leo T is the only dwarf spheroidal with gas; it is marked as a light blue star.
Red triangles are the M31 dwarfs \citep{jason}.
\label{MbV}}
\end{figure*}

Provided with estimates of the baryonic masses and circular velocities of the 
Local Group dwarf satellites, we can include them in the BTFR.  Fig.~\ref{MbV}
shows the BTFR with the Local Group dwarfs together with bright spirals \citep{M05}
and gas rich, late type disks \citep{stark,trach}.  The latter provide a calibration
of the BTFR that is largely independent of the IMF \citep{stark}.
The best fit slope is not meaningfully different from 4, so fixing the slope to this value and
combining the gas rich galaxy data of \citet{stark} and \citet{trach} leads to
\begin{equation}
M_b = A V_c^4
\label{BTFR}
\end{equation}
with $A = 45 \pm 10 \;\mathrm{M}_{\sun}\,\mathrm{km}^{-4}\,\mathrm{s}^4$ \citep{M10}.
Fig.~\ref{MbV} also shows the residuals after dividing out equation~(\ref{BTFR}).  
The residual $F_b = M_b/(AV_c^4)$ is the fraction of baryons a galaxy has
relative to the expectation of the BTFR.  

The Local Group dwarfs diverge systematically from the BTFR.
Figs.~\ref{resid_A} and \ref{resid_B} shows their residuals as functions of various quantities.  
Correlations are apparent with both luminosity and ellipticity, with fainter and less round dwarfs lying
farther from the BTFR established for disk galaxies.  
The residuals also correlate with size, surface brightness, and metallicity.
In these cases, the correlation is not as strong as with luminosity, and appears simply to
follow from the fact that these quantities correlate with luminosity themselves.  
The residuals correlate only weakly with current galactocentric distance, though there is
a clear trend among the ultrafaint dwarfs of the Milky Way.  This suggests a possible
role for the orbits of the dwarfs.  

\placefigure{MbV}

\citet{oldbellazzini} found a correlation of the surface brightnesses of dSph satellites with the 
combination of luminosity and absolute magnitude $|M_V|+6.4 \log(D)$.  They argued that this 
suggested a role for environmental influences in shaping the structure of the dwarfs.  
We find that this combination of luminosity and absolute magnitude also correlates with
the degree of deviation form the BTFR (Fig.~\ref{resid_B}).  Indeed, it extends to include the ultrafaint dwarfs,  
which of course were not know at the time that \citet{oldbellazzini} identified this quantity as relevant.

\citet{oldbellazzini} further define a dimensionless tidal force $F_{T,D}$ as an indicator of the potential
importance of the gravity of the host galaxy on its satellites.  For spherical masses in circular orbits,
\begin{equation}
F_{T,D} = \frac{M}{m}\left(\frac{r}{D}\right)^3.
\label{susceptible}
\end{equation}
Other mass distributions follow the same dimensional behavior.  
In equation~(\ref{susceptible}), $m$ is the mass of the satellite and $r$ is its size while 
$M$ is the mass of the host and $D$ is the radius of the orbit.  In Fig.~\ref{resid_B}
we use the deprojected 3D half light radius $r_{1/2}$ as a measure of
each dwarf's size, and set $m = M_b$ and $M_{host} = V^2D/G$ where $V$ is the orbital velocity
at the present location of the satellite.  For the Milky Way we use the model rotation curve of \citet{mcgMW}
for the orbital velocity, which is typically $\sim 190\;\mathrm{km}\,\mathrm{s}^{-1}$ at the distance of
most of the dwarfs \citep[see also][]{XueMW}.  For M31 we use 
$V_{\mathrm{M31}} = 250\;\mathrm{km}\,\mathrm{s}^{-1}$ \citep{M31RC}. 
The larger $F_{T,D}$, the more susceptible a satellite is to
the gravitational influence of its host.  We find that the deviation from the BTFR correlates with 
$F_{T,D}$, a point we examine in greater detail in \S \S~\ref{starstrip} and \ref{MOND}.

\subsection{Accuracy}
\label{accuracy}
 
At first glance, the dwarfs follow the same trend as the spirals, just with larger scatter.  
Our first concern is therefore whether the data are accurate enough to falsify the null
hypothesis that all of the dwarf Spheroidals are consistent with the BTFR.
The classical dwarfs, for which the data are best,
adhere fairly well to the BTFR.  Indeed, Carina, Fornax, Leo I, Leo II, Sculptor, and Sextans
follow the BTFR as well as rotating disks do.  

The M31 dwarfs also lie near to the BTFR, albeit with larger scatter.
Three (And I, X, and XIV) are consistent with the BTFR
while the other three from the \citet{jason} sample (And II, III, and VII) sit somewhat above it.
This might happen if equation~(\ref{beta}) understates the circular velocity.
The kinematic data we adopt for the M31 dwarfs are very recent.  If we look back to a time
when the data for the classical dwarfs were in a similarly early state, their scatter goes up
\citep{GS92,milg7dw}.  We therefore consider the M31 satellites to be broadly consistent 
with the BTFR, given the uncertainties.  And II is the most deviant case, being about $2 \sigma$ away. 
This is the situation for the most recent measurement, 
$\sigma_{los} = 7.3 \pm 0.8\;\mathrm{km}\,\mathrm{s}^{-1}$ \citep{jason}.  
This case illustrates the modest yet important way in which the data can evolve. 
If we adopt $\sigma_{los} = 9.3 \pm 2.7\;\mathrm{km}\,\mathrm{s}^{-1}$ as measured by
\citet{andiicote}, no discrepancy with the BTFR is inferred.

The majority of the MW ultrafaint dwarfs, together with Draco and Ursa Minor, 
lie systematically below the BTFR.  Not all of the recently discovered dwarfs deviate.
Canes Venatici I, Leo T, and Leo IV are all consistent with the BTFR within the errors.
However, Bootes I, Canes Venatici II, Coma Berenices, Segue 1, Ursa Major I, Ursa Major II and Willman 1
all deviate substantially and significantly ($> 2\sigma$) from the BTFR.  This corresponds
to deviation by at least an order of magnitude ($F_b^{-1} > 10$). 
If a population of as-yet undiscovered ultrafaint dSphs exist with large dispersions
and extremely low surface brightnesses \citep{mia,stealth}, we expect them to 
deviate still farther from the BTFR.

Hercules is an interesting case.  It either deviates or it does not, depending on which published 
velocity dispersion we adopt.  For consistency with \citet{boom}, we use here the velocity dispersion 
$\sigma_{los} = 5.1 \pm 0.9\;\mathrm{km}\,\mathrm{s}^{-1}$ found by \citet{SG}.  
For this $\sigma_{los}$, Hercules deviates substantially from the BTFR, 
being off by a factor ${F_b}^{-1} \approx 25$ in mass, with a 68\% confidence interval
$7 < F_b^{-1} < 45$.  Put this way, it would seem safe to say that it deviates by an order of magnitude.  
However, if instead we adopt the more recent 
measurement\footnote{The difference in the velocity dispersions stems
from how members of the system are identified, a systematic effect that is not represented by
the formal uncertainty \citep[see also][]{serradw}.} of 
$\sigma_{los} = 3.7 \pm 0.9\;\mathrm{km}\,\mathrm{s}^{-1}$ reported 
by \citet{aden}, we find $F_b^{-1} \approx 7$ with a 68\% confidence interval
$0.2 < F_b^{-1} < 13$.  The central value again deviates by a large factor, and
the uncertainties in the two measurements overlap.  However, the $\pm 1\sigma$ boundaries
now encompass the BTFR, so we would no longer consider this to be a deviant case. 

\begin{figure*}
\epsscale{1.0}
\plotone{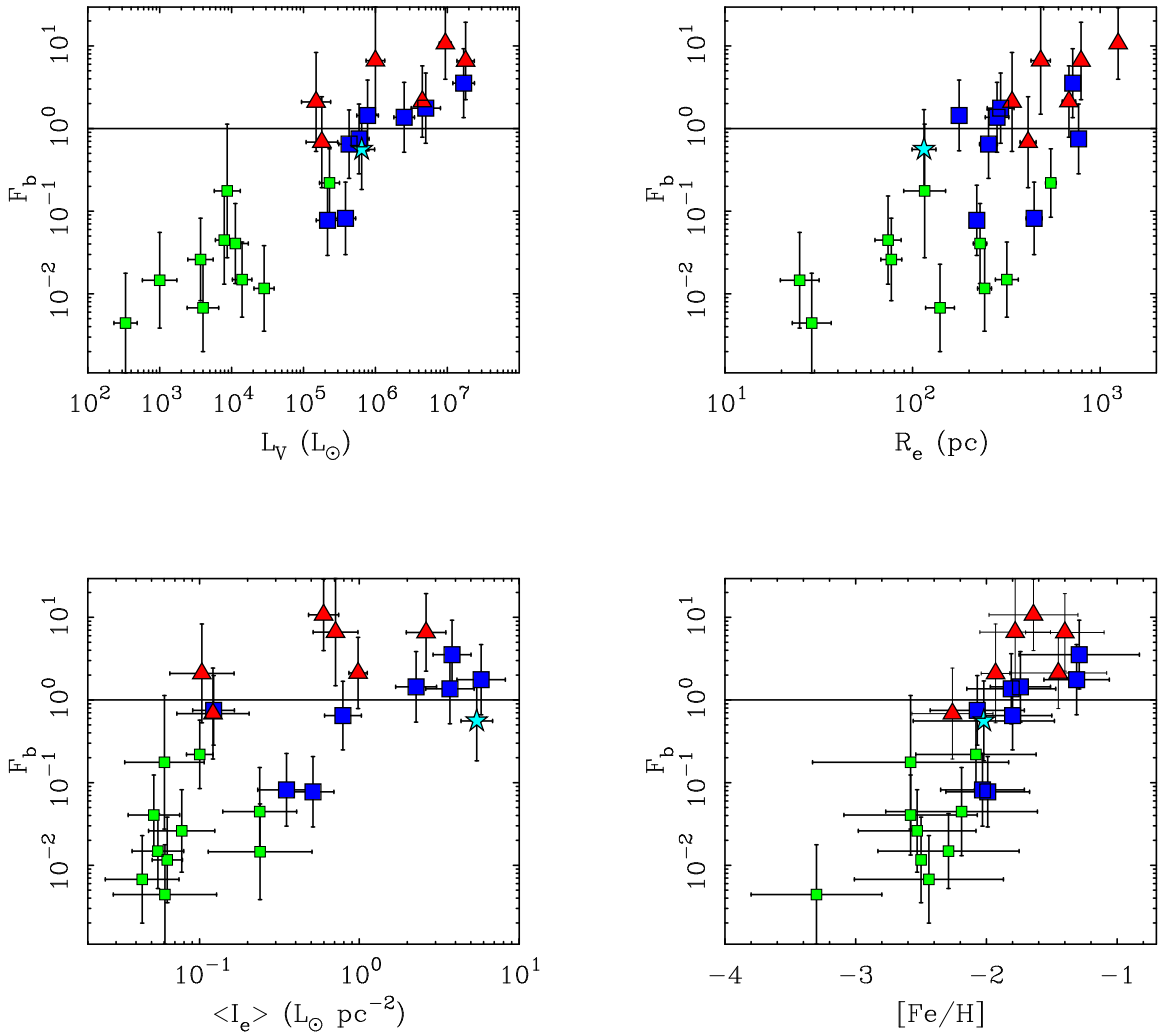}
\caption{Residuals from the BTFR correlate with luminosity (top left), size (top right),
surface brightness (bottom left), and metallicity (bottom right) in the sense that the faintest,
smallest, dimmest, and most metal poor dwarfs deviate farthest from the BTFR.
Symbols as per Fig.~\ref{MbV}.
\label{resid_A}}
\end{figure*}

\placefigure{resid_A}

The case of Hercules illustrates an important point.  A small and formally consistent difference
($1.4\pm1.3\;\mathrm{km}\,\mathrm{s}^{-1}$ in this case) can have an impact on how we perceive the
the results under consideration here.  It is therefore important to take the formally
significant deviations with a grain of salt.  This is a rapidly evolving field; even with heroic efforts
the velocity dispersion data for the ultrafaint dwarfs are not of the same quality as those for the classical
dwarfs, nor can they be given the small number of stars in these systems.

\begin{figure*}
\epsscale{1.0}
\plotone{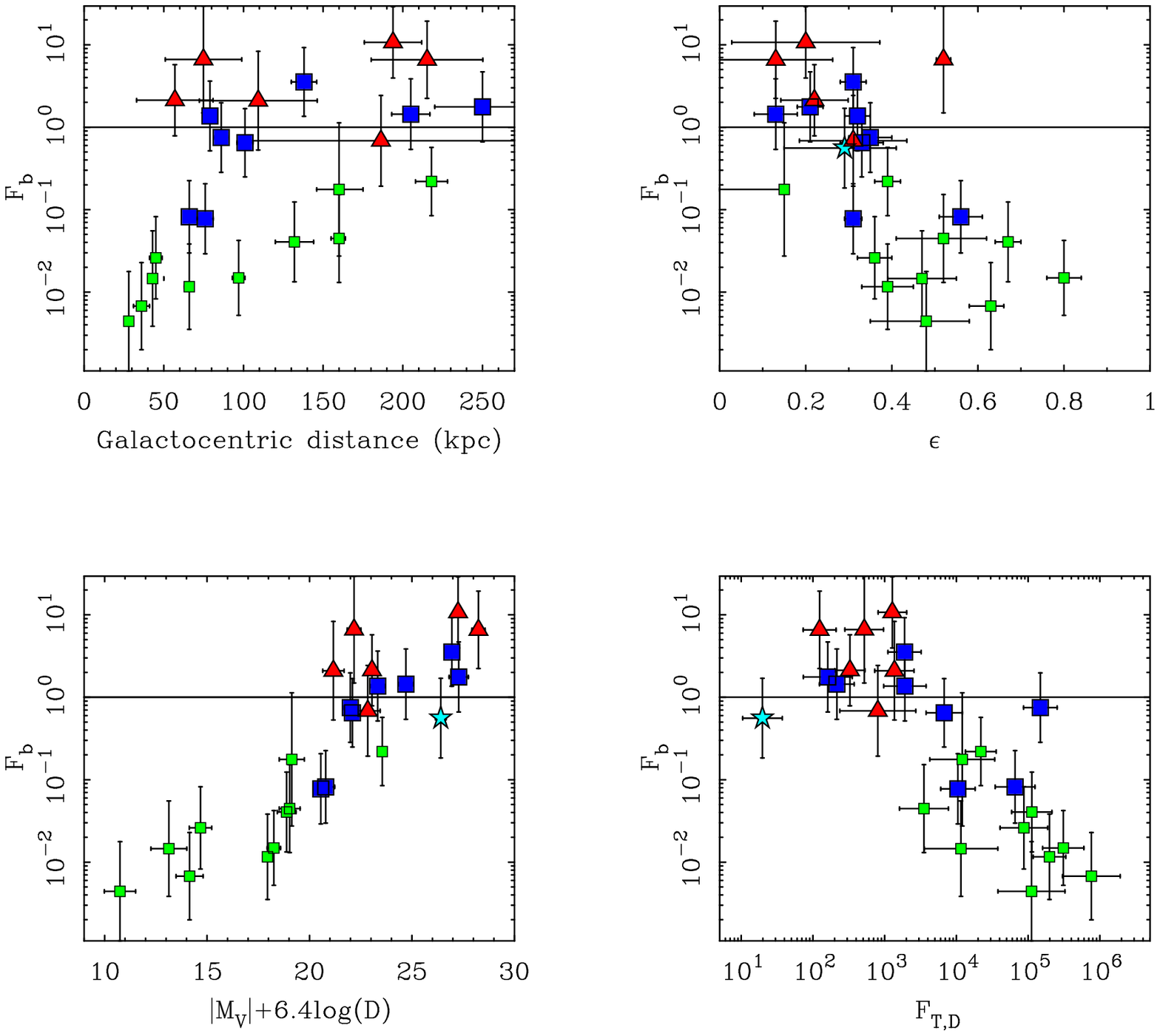}
\caption{Residuals from the BTFR correlate with indicators of environmental and
and tidal influence such as distance from the host galaxy (top left), shape (as measured by
the projected ellipticity of the dwarf, top right), the combination of luminosity and distance
found by \citet[bottom left]{oldbellazzini}, and a 
measure of the dwarfs' susceptibility to tidal influences (bottom right; equation \ref{susceptible}).
The most deviant objects tend to be the closest (especially among the ultrafaint dwarfs) 
and least spherical.  Leo T falls off the right edge of the top left plot with $D = 407$ kpc.
Symbols as per Fig.~\ref{MbV}.
\label{resid_B}}
\end{figure*}

\placefigure{resid_B}

Given the cautionary tale of Hercules,
one possible interpretation is that there are no real deviations from the BTFR.  
This hypothesis predicts that the scatter will decrease as the data improve,
with the more deviant cases migrating towards the BTFR.  There is room to improve the data 
by weeding out nonmember stars \citep{serradw} and binaries \citep{minordw}, both of which
inflate $\sigma_{los}$.  Measuring proper motions to constrain anisotropies will give a more
accurate assessment of $V_c$ \citep[e.g.,][]{strig07}.  Of course, we must also consider the degree to which it is 
appropriate to compare $\sigma_{los}$ measured at small radii 
in dwarfs to $V_c$ measured at large radii in disk galaxies.

There are substantial difficulties with attributing all deviations from the BTFR to errors.
One is their sheer magnitude:  the discrepancies $F_b$ are an order of magnitude in some cases,
and two orders of magnitude for some of the ultrafaint dwarfs.  In the most extreme case, Segue 1,
$F_b^{-1} \approx 230$.  To place this object on the BTFR would require that its baryonic mass
increase by this factor.  Alternatively, since the BTFR is steep, a reduction in the 
observed $\sigma = 4.3 \pm 1.1\;\textrm{km}\,\textrm{s}^{-1}$ \citep{geha09} to
$\sigma_{los} = 1.1\;\textrm{km}\,\textrm{s}^{-1}$ would also suffice.  While a $3 \sigma$
change may be conceivable in one object, it would have to happen in numerous cases.
Moreover, the residuals are asymmetric:  many more systems have $F_b < 1$ than $F_b > 1$.
This asymmetry is larger than indicated by the formal errors, though we caution that the errors
are asymmetric in the same sense.  As always,
a systematic error could be responsible, such as the inflation of $\sigma_{los}$ by binaries.
Any such systematic would have to similarly affect many different objects observed by a 
number of independent groups while not affecting the objects that do adhere to the BTFR.  

Another problem with attributing deviations from the BTFR entirely to uncertainties is that the 
residuals correlate with the physical properties of the dwarfs (Figs.~\ref{resid_A} and \ref{resid_B}).  
It seems unlikely that the observed correlations are mere flukes.  
If not, they will persist as data continue to accumulate.

\subsection{Gas Removal}

If the BTFR is a property shared by isolated galaxies when initially formed, we might seek to
reconcile the Local Group satellites with it by extracting baryons before they form stars. 
Mechanisms to do this come in a variety of flavors.  In this section we consider astrophysical
effects that systematically suppress the cooling or retention of gas in dwarfs.

\subsubsection{Cosmic Reionization}

The reionization of the universe at $z > 6$ is one mechanism by which the formation of stars
in low mass halos might be suppressed.  Reionization heats the gas, inhibiting
further star formation in small halos
\citep[$V_c \lesssim 20\;\mathrm{km}\,\mathrm{s}^{-1}$ at the time of reinoization: see, e.g.,][]{bullockreion,krav}.  
Halos that remain this small may never experience further star formation, persisting as 
``fossil'' dwarfs containing only ancient stars \citep{ricotti,mia}.  

\begin{figure}
\epsscale{1.0}
\plotone{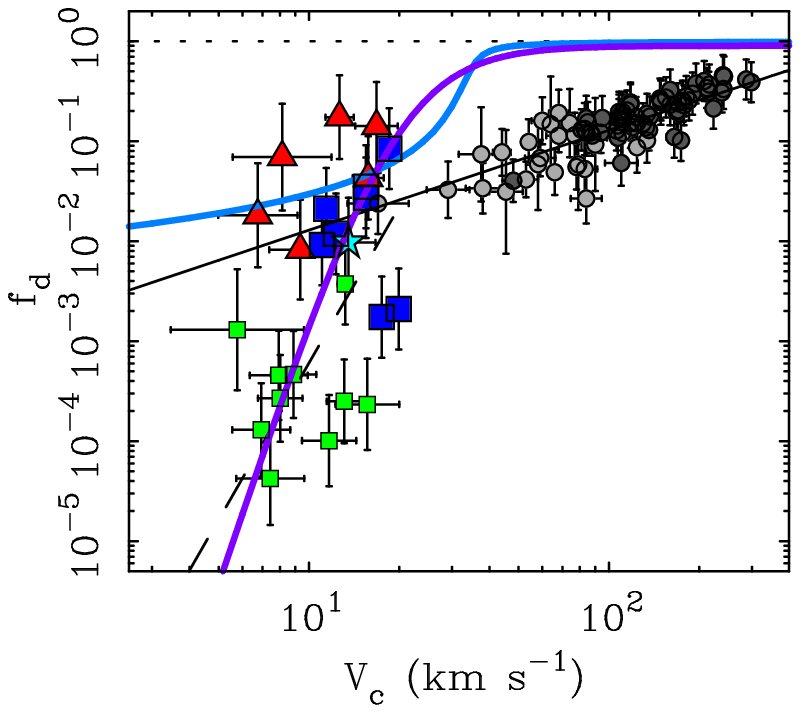}
\caption{The detected baryon fractions $f_d = M_b/(f_b M_{500})$ \citep{M10} of galaxies 
(symbols as per Fig.~\ref{MbV}) fall well below the cosmic baryon fraction $f_b$ (dotted line).  The 
ultrafaint dwarfs appear to follow a steeper relation than the BTFR (thin solid line):
the dashed line is the fit from \citet{M10} before the M31 data \citep{jason} were available.
The heavy solid lines illustrate the expected effects of cosmic reionization 
\citep[descending line:][S-shaped line]{crain,hoeft}.  
This alone cannot simultaneously explain both dwarfs and disks.
\label{fdVLG}}
\end{figure}

\placefigure{fdVLG}

The results of simulations \citep{crain,hoeft}
are compared to the data in Fig.~\ref{fdVLG}.  Here we plot the detected baryon fraction,
which is the fraction of baryons $f_d = M_b/(f_b M_{500})$
that are detected in any given halo relative to the cosmic
baryon fraction $f_b$ \citep{M10}.  Note that $f_d$ differs from
the BTFR residual $F_b$.

Reionization models like those of \citet{crain} and \citet{hoeft} result in a sharp truncation of the
cold baryon content of halos at a particular mass scale.  Above this mass scale, reionization
does nothing to prevent the baryons from cooling and forming stars.  In contrast, 
the data deviate smoothly and gradually from the universal baryon fraction, not abruptly
as suggested by models consisting of reionization only.  Thus, while the model of \citet{crain} appears
to work well for $V_c < 20\;\mathrm{km}\,\mathrm{s}^{-1}$, it does not explain
dwarf Irregulars with $V_c \gtrsim 30 \;\mathrm{km}\,\mathrm{s}^{-1}$.  
At this slightly larger scale, reionization is not
expected to impede the condensation of cool gas and subsequent star formation, yet the mass of baryons
detected in the form of stars and cold gas is an order of magnitude shy of the cosmic baryon fraction.  

We might suppose that there is some fundamental difference between dSph and dIrr galaxies that
separately determines their location in Fig.~\ref{fdVLG}.  However, the continuity of the BTFR to most
of the classical dwarf spheroidals suggests a common origin.  
It appears that physics beyond reionization is needed to explain the observed
trend in detected baryon fraction $f_d$ with halo mass.  

Reionization might act in addition to whatever physics drives the main trend in Fig.~\ref{fdVLG} 
by causing a sharp low-velocity\footnote{An interesting test of the idea that 
reionization imposes a sharp cut-off to the existence of
low mass galaxies would be if there exist any galaxies of exceedingly low rotation velocity
($V_c \lesssim 5\;\mathrm{km}\,\mathrm{s}^{-1}$) that  persist in following the BTFR,
especially if they were gas rich.} cut-off.  
This might explain the residual correlation with luminosity exhibited by the ultrafaint dwarfs 
(at least those with $L_V < 10^5\;\textrm{L}_{\sun}$ in Fig.~\ref{resid_A}).  
However, it does nothing to explain the trend of the residuals with ellipticity (Fig.~\ref{resid_B}).

\subsubsection{Stellar Feedback}

Stellar feedback is frequently invoked as a mechanism to remove gas from galaxies or 
inhibit its accretion in the first place.
Intense radiation from young stars \citep[perhaps Pop.~III stars at early times, e.g.,][]{RO04}
may heat the surrounding gas and prevent its infall. 
Winds driven by supernovae following intense bursts of star formation may impart
sufficient mechanical energy into the interstellar medium to sweep up and expel substantial
amounts of gas \citep[e.g.,][]{MLF99,GZ02}.  In either case, the prodigious energy produced by massive, 
short-lived stars is invoked to alter the global properties of galaxies.

These ideas go back to at least \citet{DS} and have been invoked many times
\citep[e.g.,][]{MF,priya,efst,vdB00,MayerMoore,ricotti,monstermash,aaron,fabio,sawala}.  
One appealing aspect of this picture is that the energy injected by star formation 
events depends on the local star formation efficiency, while the ability of a halo to retain
baryons depends on the depth of its potential well.  It is therefore natural to expect
a trend like that seen in Fig.~\ref{fdVLG} with the retained baryon fraction
increasing with halo mass.  Star formation happens in a local way, but global baryon
retention becomes increasingly difficult in smaller galaxies.  When the potential well
becomes very shallow, as is the case for dwarf spheroidals, stochastic effects may
come to dominate completely such that $F_b$ varies widely at a given $V_c$.

While qualitatively attractive, quantitative implementation of feedback in realistic simulations is 
notoriously difficult \citep[e.g.,][]{fabio}.  The gas physics is uncertain, and even the depth
of the potential well is unclear.  In the \LCDM\ structure formation paradigm, 
large galaxies are constructed from the mergers of smaller ones.  The trend of $f_d$ with 
potential well depth only follows naturally if most of the star formation activity responsible for the feedback 
takes place after the establishment of the bulk of the potential well depth
as indicated by the circular velocity measured now.  If instead much of the action occurs in small
protogalactic clumps before the establishment of the large halos we imagine
to retain the gas, it becomes less obvious that the ultimate potential well depth is relevant.
Indeed, as our nearest examples of the fossils of protogalactic clumps,
the large scatter in luminosity and $f_d$ at a given mass exemplified by the 
dwarf spheroidals more nearly represents the stochastic mess that 
we would naturally expect from this process.

Given that stellar feedback is a stochastic process, any relation that it establishes will likely
have a great deal of scatter.  In the context of the Local Group dwarfs, this is quite
reasonable:  there is a lot of scatter about the BTFR (Fig.~\ref{MbV}).  This does not persist
to larger masses ($V_c > 30\;\mathrm{km}\,\mathrm{s}^{-1}$).  For the gas dominated late
type disks populating Fig.~\ref{fdVLG} \citep{stark,trach}, the data are consistent with a relation
with nearly zero intrinsic width.  These data already place an uncomfortably tight limit on the 
triaxiality of dark matter halos \citep{trach,KdN09}; there is not much room for additional stochastic effects. 

Note also that rotating galaxies adhere to the BTFR regardless of their gas 
fraction \citep{andersonbregman}.
This is contrary to the natural expectation of feedback scenarios in which 
galaxies that have converted more gas into stars have
experienced more star formation and hence more feedback.
For the dwarf spheroidals, the increased scatter about the BTFR
could be the signature of a stochastic effect, if not for the fact that
the residuals are not random as they correlate with other physical properties
\citep[Figs.~\ref{resid_A} and \ref{resid_B}; see also][]{walkerandme}.

Recent simulations \citep[e.g.,][]{stinson, salvadori,sawala} of low mass galaxies 
appear to do a reasonable job of reproducing some of their observed properties.  It will be
interesting to see if they are equally successful in matching the correlations of residuals from
the BTFR.  The correlations with luminosity and metallicity may indeed follow, 
as the chaos of feedback can easily lead to a wide range of luminosity at a given halo mass, and
supernova winds may carry away much of the metals they produce.  The
correlation with ellipticity and tidal susceptibility is less obviously explicable in this way.

\subsubsection{Ram Pressure Stripping}
\label{rampress}

Another mechanism capable of removing cold gas before it can form stars is ram pressure stripping.
In this hypothesis, the initial condition is gas rich galaxies that adhere to the BTFR.  
As these objects fall into larger structures like the halo of the Milky Way, hot gas in the host's halo
ablates the cold gas of the infalling satellites.  The resulting object is under-luminous with respect 
to the BTFR because many baryons were removed before they could form stars.

The Magellanic stream may provide an example of this process in progress \citep{mastropietro},
though tidal effects might also suffice \citep{MStideC,MStideS}.  Certainly it is striking that all
of the nearby ($D < 250$ kpc) dwarf spheroidals are devoid of gas \citep[e.g.,][]{LeoTgas}.
Provided that there is sufficient hot gas in the halo, ram pressure stripping should occur.

In this hypothesis, we associate the stripping of the cold
gas baryons with the epoch when the object fell into the larger structure.  
The galaxies that fell in first would have had the least opportunity to form stars
and would have lost the most gas.  This has two potentially testable consequences.
First, the amount by which a dwarf deviates from the BTFR is an indicator of the infall epoch.
The earlier it fell in, the sooner gas was stripped and star formation truncated.  
Secondly, because of the truncation of star formation, the age of the stars should also 
correlate with the factor by which the dwarf deviates.  Dwarfs that fell in early should
contain only old stars.  A natural corollary is that the stars in the first dwarfs to fall in
should be low metallicity.  Indeed, since these objects would have had little time to self-enrich,
we would expect the [$\alpha$/Fe] ratios to be enhanced while [Fe/H] is low.  We would
thus expect [$\alpha$/Fe] to correlate with the residuals $F_b$.

The degree of deviance $F_b$ should reflect the order of infall. 
The dwarfs that deviate most presumably do so because they fell in earliest.  If this holds, then
Segue 1 and Ursa Major II fell in first, followed by Willman 1, Bo\"otes I, Ursa Major I,
and Coma Berenices.  Hercules, Canes Venatici, Draco, and Ursa Minor are relative newcomers,
while Leo T, which is the only MW dSph with detectable HI gas, has yet to be stripped.

The prediction that deviance correlates with infall time is not perfect, as it depends on the
star formation history of the dwarf prior to infall.  An object that converted most of its cold
gas into stars quickly before infall would not deviate much from the BTFR even if the infall
were relatively early.  However, no star formation can persist after ram pressure stripping is
complete.  We do therefore expect the dwarfs composed of the oldest stars
to be the first to have fallen in.  

An interesting corollary has to do with the epoch of most recent star formation.
Some of the classical dwarfs have had complex star forming histories \citep{grebel},
and are not composed solely of ancient stars.  In some cases, the last episode of
star formation is rather recent \citep[e.g.,][]{grebel,HGV,dolphin,helmi,martinLBT,THT}.  
This has been a puzzle, since, with the exception of
Leo T, none of these objects contain substantial quantities of cold gas at present.
The truncation of star formation by infall predicted by the ram pressure stripping hypothesis
provides a natural explanation for this puzzle.  The dwarfs with recent star formation episodes are
merely the most recent to have fallen in and had their gas stripped away.

For this process to be generic, the halos of host galaxies need to contain a good deal of hot gas.
\citet{LeoTgas} infer a density of hot gas $n > 2 \times 10^{-4}\;\textrm{cm}^{-3}$ 
out to at least 70 kpc around the Milky Way in order for ram pressure stripping to have effectively
ablated the local population of dwarf spheroidals.  In contrast, \citet{andersonbregman} place
a limit $n < 7 \times 10^{-5}\;\textrm{cm}^{-3}$ at 50 kpc based on pulsar dispersion measures
towards the LMC and other methods.  This would not be enough for ram pressure stripping to 
have played a major role in shaping the history of satellite dwarfs.  The mass and extent
of the hot gas halos around spiral galaxies is an important but ill-constrained piece of the puzzle.

The effects of ram pressure stripping are not limited to the small scales of interest here.
It is observed to affect spiral galaxies falling into rich clusters \citep[e.g.,][]{aeree}.
If this mechanism causes dwarf spheroidals to deviate from the BTFR in the Local Group,
we might expect it to have the same effect on spirals in clusters.  However, this is not observed 
\citep{GvGKB,aeree}.  Since hot gas is definitely present in rich clusters, and cold gas is
observed to be stripped from spirals there, one might expect the effect to be more pronounced
in clusters rather than less.  However, stellar mass dominates the baryon content of the 
infalling spirals, which typically have modest gas fractions ($\sim 0.1$ -- 0.2).  This might
suffice to mask the effect, though an investigation of the impact of different environments
on the BTFR would clearly be interesting.

Ram pressure stripping potentially provides a good way to strip gas at an appropriate time
to allow for recent star formation in some dwarf spheroidals that possess no cold gas now.
It would naturally account for the correlation of BTFR residuals with luminosity and metallicity
if enough gas is stripped.
It does not, by itself, provide an obvious explanation for the correlation with ellipticity and
tidal susceptibility.  These are features of the stellar distribution that are difficult to explain
with any mechanism that acts solely on the gas.  

\subsection{Stellar Stripping}
\label{starstrip}

Tides are a mechanism that acts on the stars directly.
The correlation of BTFR residuals with ellipticity and tidal susceptibility
as well as with luminosity would occur naturally if some of a dwarf's original stars have been 
stripped away by tidal perturbations as it orbits the Milky Way.  
Under this hypothesis, the deviant dwarfs originally adhered to
the BTFR, but have had much of their original luminosity removed.
Much of the stellar halo of the Milky Way might have been built up this way 
through hierarchical merging \citep{kathryn,bj05}.

\begin{figure}
\epsscale{1.0}
\plotone{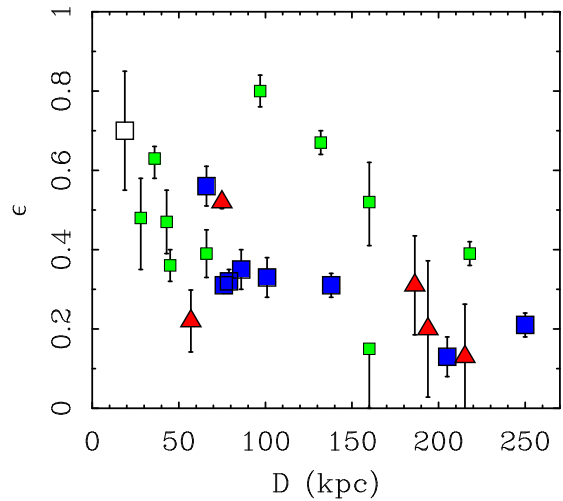}
\caption{The ellipticity of Local Group dwarf Spheroidals as a function of distance from their host galaxy.  
Symbols as per Fig.~\ref{MbV}.  The open square is Sagittarius \citep{ibata}.  
Leo T ($\epsilon = 0.29$ at $D = 407$ kpc) falls off the right edge of the plot.
%The shapes of satellites become less round as they approach their host, suggesting a role for tides.
\label{ellD}}
\end{figure}

\placefigure{ellD}

Analyses of the dynamics of the dwarf spheroidals typically assume spherical symmetry.  
The large ellipticities of many of the dwarfs call this assumption into question.  
It is manifestly untrue in Canes Venatici II, Ursa Minor, Ursa Major I and II, and Hercules, 
all of which have $\epsilon \gtrsim 0.5$.  In this context, it is important to determine whether 
some dwarfs show dynamical evidence of disruption \citep[e.g.,][]{munozcar,munozUMII}.

The fact that the degree of deviation from the BTFR 
correlates with ellipticity (Fig.~\ref{resid_B}) may be a clue that tidal disruption is playing a role.  
A further clue is provided by Fig.~\ref{ellD}, which shows that the ellipticity of the dwarfs also correlates
with distance from their host.  While we do not expect all dwarf Spheroidals to be perfectly round,
neither do we expect their intrinsic shape to depend on distance unless some environmental factor
is playing a role \citep[see also][]{oldbellazzini}.  

If ellipticity is an indicator of tidal disruption then the correlation of deviations with luminosity follows 
naturally \citep[cf.][]{piatektides,munozcar}.
The more disrupted a galaxy, the more elliptical it becomes, and the more stars it has lost.
The more stars a dwarf has lost, the further it deviates from the BTFR.
An obvious observational test of this hypothesis would be to search for streams associated 
with the deviant dwarfs \citep[e.g.,][]{munozUMi,fellhauer,salesorphan,newbergorphan}.
The age and metallicity of stars in these streams should be consistent
with having been drawn from the distribution present in the parent body.  

Sagittarius is one example where this has clearly happened.  
It is now out of equilibrium, so we would not expect it to reside on the BTFR.
If we adopt a luminosity of $\sim 2 \times 10^7\;\textrm{L}_{\sun}$ \citep{mateo96},
the BTFR predicts $\sigma_{los} \sim 16\;\textrm{km}\,\textrm{s}^{-1}$, somewhat larger than
the observed $\sim 11\;\textrm{km}\,\textrm{s}^{-1}$ \citep{ibata,bellazzini}.
Given the uncertainties, it is not obvious that this constitutes a significant discrepancy.
However, if the larger total luminosity estimate ($\sim 10^8\;\textrm{L}_{\sun}$) of
\citet{NewSag} is correct, then Sagittarius should initially have had 
$\sigma_{los} \sim 24\;\textrm{km}\,\textrm{s}^{-1}$.
We need to understand how luminosity and velocity dispersion evolve as
disruption takes place.  

A satellite in orbit around a much larger
parent galaxy is obviously subject to tidal disruption \citep{kathryn}, but it is less obvious how
$\sigma_{los}$ will evolve as mass is stripped \citep[see, e.g.,][]{stripklim,striplokas}.  
Naively, we would expect the initial stripping
to primarily affect the outer dark halo first.  Stars would not be stripped in significant numbers until
much of the dark halo was already gone \citep{pendwA}.  
%If the recurring inference of a minimum halo mass \citep{mateo,gilmore,strigari} is correct, none of the dwarfs considered here would have lost enough halo mass to expose the stars to stripping.  Of course, this inference assumes equilibrium dynamics, which might not hold.  

We can be more quantitative about the propensity of a dwarf to be stripped by returning
to the subject of  tidal susceptibility (equation~\ref{susceptible}).  It is conventional to define
the tidal radius \citep{keenanb,keenana} as
\begin{equation}
r_{t,D}  = D \left(\frac{m}{3M}\right)^{1/3}.
\label{RTconventional}
\end{equation}
This is approximately the radius where a star contained in a satellite of mass $m$ is subject
to being lost to the host mass $M$, though the details depend on the orbit.  
Equation~(\ref{RTconventional}) implicitly assumes circular orbits; eccentric orbits
are most susceptible to tidal stripping near pericenter where 
$D \rightarrow \mathrm{a}(1-\mathrm{e})$ (where $\mathrm{a}$ is
the semi-major axis and $\mathrm{e}$ is the eccentricity of the orbit).  
Equation~(\ref{RTconventional}) provides an estimate for when stripping may occur,
bearing in mind that the current galactocentric distances of the dwarfs could be larger
than their pericenter distances.

\begin{figure*}
\epsscale{1.0}
\plotone{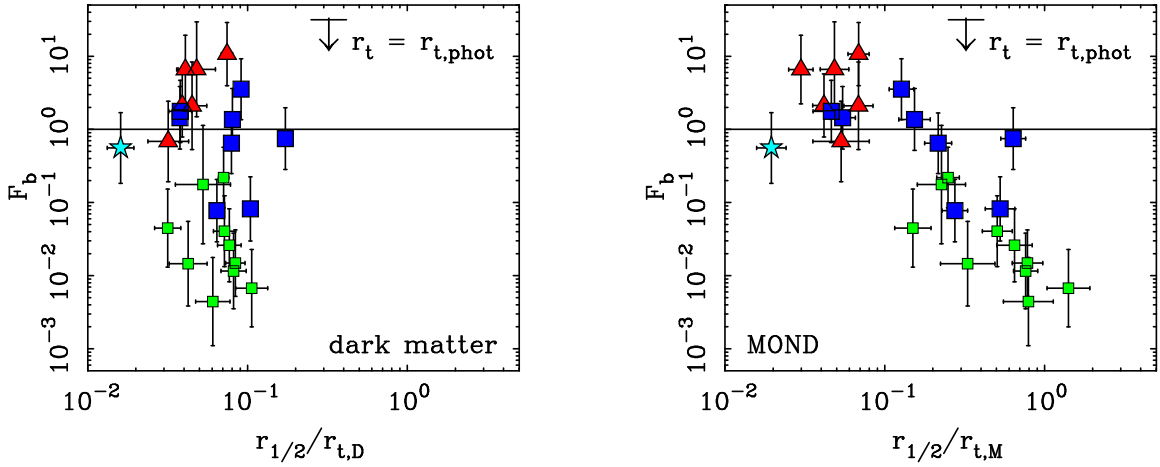}
\caption{The residuals from the BTFR (symbols as per Fig.~\ref{MbV})
plotted against the ratio of half-mass radius to tidal
radius in the case of dark matter (left) and MOND (right).  The plots are scaled identically.
The tidal radius depends on the masses of satellite and host (equations \ref{RTconventional}
and \ref{RTMOND}).  Masses are computed dynamically for the case of dark matter 
(left: $m = M_{1/2}$) and using only baryonic mass in the case of MOND (right: $m = M_b$).  
The arrow marks the location where the average photometric tidal radius equals the computed
tidal radius.  In the case of MOND, the location where dwarfs deviate form the BTFR corresponds 
well to the observed photometric tidal radius, and the amount of deviation correlates with the size of a
dwarf relative to its MONDian tidal radius.
\label{tidalR}}
\end{figure*}

Fig.~\ref{tidalR} shows the sizes of the dwarfs in terms of tidal radii.  For a star orbiting at the 
deprojected 3D half light radius, we set $m = M_{1/2} = 3 \sigma_{los}^2 r_{1/2}/G$ \citep{boom}.  
The distance $D$ is the galactocentric distance of each dwarf from its host, and the mass
of the host is the mass contained within $D$:  $M = V^2 D/G$.  For the Milky Way we
use $V_{\mathrm{MW}}(D)$ from \citet{mcgMW} while for M31 we assume that a constant velocity suffices at
the location of the M31 dwarfs and adopt $V_{\mathrm{M31}} = 250\;\mathrm{km}\,\mathrm{s}^{-1}$
\citep{M31RC}. 

\placefigure{tidalR}

The tidal radii of the dwarfs are typically an order of magnitude or more larger than their half light radii.
By this criterion, the stars of all the dwarfs are safely cocooned deep inside their dark matter halos
and are in no danger of being stripped \citep{pendwB}.  The vast majority of the dark matter must
be tidally stripped before the stars become exposed to stripping \citep{pendwA}.  It therefore appears
that these dwarfs should not currently be losing stars.  This is somewhat surprising, given the large
deficit $F_b$ of some of these dwarfs, and the correlation with tidal susceptibility seen in
Fig.~\ref{resid_B}.

Photometric tidal radii are less certain than the deprojected 3D half-light radius plotted in Fig.~\ref{tidalR},
but correlate very well with it.  On average, $r_{t,phot} = (3.1 \pm 0.8) r_{1/2}$ for the dwarfs with
photometrically well measured tidal radii.  We mark this location in Fig.~\ref{tidalR}.  
The mean size of the dwarfs is only $r_{1/2} = 0.06 r_{t,D}$.  To bring the 
photometric tidal radii into agreement with the dynamical tidal radii computed with 
equation~(\ref{RTconventional}) would require an adjustment to the mass of nearly
two orders of magnitude:  $m \approx 0.01 M_{1/2}$.  This seems like a lot to ask, but there are
several possibilities.

First, $M_{1/2}$ may be overestimated because the disrupting systems are not in equilibrium.
The true mass might be less, allowing tidal escape.  In this case the velocity dispersion is artificially
inflated by escaping stars.  Though qualitatively attractive, this effect is probably not large enough.
From numerical simulations, \citet{striplokas} estimate that the inflation of the velocity dispersion due to 
stripping should result in an overestimate in mass of at most 60\%.  Nevertheless, one can imagine
that the amount of dark matter is less than it might seem.

In this context, it is worth noting that at least some dark matter is required.
Equation~(\ref{RTconventional}) differs from equation~(\ref{susceptible}) only by a factor of $3^{1/3}$, 
yet the result in the left panel of Fig.~\ref{tidalR} differs from that in the lower right panel of
Fig.~\ref{resid_B} because there we set $m = M_b$.  
Dwarfs lacking in dark matter are quite susceptible to tidal influences
\citep{kroupadwarf}.  Indeed, using the baryonic mass $M_b$ in place of the dynamical
mass $M_{1/2}$ in equation~(\ref{RTconventional}) results in tidal radii that are
much smaller than the observed half light radii.  If this were the case, the dwarfs should have disintegrated 
long ago.

A second possibility is that $M_{1/2}$ is correct and the observed dwarfs are not yet the parents of any
streams.  The streams observed in the halo would have been produced by the dissolution of other dwarfs.
These objects dissolved in the past and are now completely gone.  While this may have happened, it
does nothing to explain the deviations $F_b$, nor their correlation with ellipticity.

A third possibility is that the orbits of the dwarfs are highly radial, a situation that is widely expected
in \LCDM\ \citep[e.g.,][]{bj05,stripklim,pendwB,striplokas,stripsales}.  If we replace 
$D$ with $\mathrm{a}(1-\mathrm{e})$, $r_{t,D}$ shrinks as $\mathrm{e} \rightarrow 1$.  
To explain Fig.~\ref{tidalR}, we need the deviant dwarfs to be on orbits with $\mathrm{e} \gtrsim 0.9$.

Most of the damage would occur during pericenter passage, when the least bound stars are lost.
The typical dwarf is not currently close to its pericenter, but may
bear the scars of its close passages in the form of reduced luminosity and deviation from the BTFR.
In this case we imagine that the streams in the halo have been built up by stars lost during successive
pericenter passages.  

This last possibility provides an appealing explanation for the correlation of $F_b$ with ellipticity as
well as luminosity.  However, we caution that the velocity dispersion is subject to evolution during this
process as well as the luminosity and shape of a dwarf.  In the simulations of \citet{pendwA}, the velocity
dispersion decreases as luminosity is lost, in a proportion that almost maintains the slope of the BTFR.
Such an evolution will not reproduce the trend observed here, which requires a more rapid loss of
luminosity with respect to velocity dispersion.  This motivates further simulation work that probes a
wider range of initial conditions and halo models. 

In the context of our empirically motivated stellar stripping hypothesis, 
we expect three broad classes of objects.  Dwarf satellites that have not yet ventured too close
to their massive host will reside on the BTFR.  The classical dwarfs (excluding Ursa Minor and Draco)
and the dwarfs of M31\footnote{Examination of the properties of the dark matter halos of the M31 dwarfs 
suggests the opposite, that they may have been more subject to tidal evolution \citep{walkerandme}.} 
fall in this category.  Dwarfs that deviate from the BTFR do so because they have
lost some stars during pericenter passages.  Greater deviation is a sign of greater damage, from either
closer pericenters or multiple pericenter passages.  Ursa Minor, Draco, and most of the MW
ultrafaint dwarfs fall in
this category.  Finally, some of the observed streams may have come from parent bodies that
have been totally disrupted by this process. 

The stellar stripping hypothesis makes several predictions.  The dwarfs that deviate from the BTFR
should be the parents of stellar streams.  These stars should
have properties consistent with the parent body: their ages and abundance patters should be consistent
with the same distributions.  The streams should be in orbits identifiable with their parents.  
It appears likely that stars would largely be liberated during pericenter passage, so there is in
principle a further prediction about the phase at which stars were injected into the stream.  
Pericenter passage is a brief portion of the orbit, so we would not expect any of the observed dwarfs to currently
be in the process of losing stars.  At their present galactocentric distances, all of the dwarfs should be 
safely bound in what remains of their dark matter cocoons.  Observation of streams being liberated at
present would require some further interpretation (\S~\ref{MOND}).

In order to liberate stars as envisioned under the stellar stripping hypothesis, 
the disrupting satellites need to be on highly elliptical orbits.
Though we do not yet know the orbits of all the dwarfs, there are some constraints.
Carina has $\mathrm{e} \approx 0.67$ \citep{piatekC} 
and Ursa Minor has $\mathrm{e} \approx 0.39$ \citep{piatekUMi}.
Carina adheres to the BTFR but Ursa Minor does not, so their observed orbits are the opposite of
what we would expect in this hypothesis.  While disfavored, a pericenter passage that is sufficiently small
($\sim 20$ kpc) to liberate some stars from Ursa Minor is not yet excluded by the orbital data \citep{LRL}.  
Even if this were the case it would remain curious that Ursa Minor deviates from the BTFR and Carina does not.  
Clearly, better orbital constraints are desirable, as are searches for liberated stars.  

\subsection{MOND}
\label{MOND}

An alternative to the \LCDM\ paradigm is the Modified Newtonian Dynamics (MOND)
hypothesized by \citet{milgrom83}.  In MOND, there is no dark matter; rather, the apparent need
for dark matter stems from a change to the force law that occurs at an acceleration scale 
$a_0 \lesssim 1\;\mathrm{\AA}\;\mathrm{s}^{-2}$.  Above this scale, all is normal:  the gravitational
acceleration can be calculated from the observed distribution of baryonic matter and Newton's
Law of Gravity:  $g = g_N$.  Below this scale ($a \ll a_0$), the modification 
$g \rightarrow \sqrt{a_0 g_N}$ applies.  This idea has its problems \citep{clowe,angusbuote}, 
but has also had more success than seems to be widely appreciated \citep{SMmond,bekenstein}.

Dwarf spheroidals provide a strong test of MOND because
their low surface densities place them deep in the modified regime.
In this regime, the BTFR is an absolute consequence of the force law for isolated objects:
$g = \sqrt{a_0 g_N}$ leads to $a_0 G M_b = V_c^4$ for point masses.
The gas dominated dwarf Irregular galaxies studied by \citet{stark} and \citet{trach} fall
along the BTFR, consistent with MOND.  Indeed, this is confirmation
of a prediction that has zero free parameters: the gas mass dominates $M_b$ and these galaxies
fall along the MOND prediction, which is indistinguishable from the best fit BTFR.   

The deviation of the Local Group dwarf satellites from the BTFR would appear to falsify MOND
\citep{GS92,milg7dw,SSH,angusdw,serradw,hernandezmond}.  However, 
the Local Group dwarf satellites are not isolated.  In MOND,
the criterion for isolation is whether the internal acceleration of an object, its self-gravity $g_{int}$, 
exceeds that imposed by external systems, $g_{ext}$.  If $g_{int} < g_{ext}$, the external field modifies
what would otherwise happen to the same object in isolation.  This external field effect is a unique feature
of MOND that has no analog in the conventional context.

Among the dwarfs considered here, Leo T is the only example that is isolated enough to be 
in the pure MOND regime such that $g_{ext} < g_{int} < a_0$.  
As such, it should obey the BTFR, and within the errors, it does. 
Sagittarius is an example that is clearly not isolated, with $g_{int} < g_{ext} < a_0$.  
This places it in the
quasi-Newtonian regime \citep{milgromQuasiN} where the mass estimator is different
\citep{milg7dw} and the BTFR is no longer absolute.  The quasi-Newtonian mass estimator 
basically just corrects the normal Newtonian mass estimator by a factor 
$G \rightarrow G_{eff} = G a_0/g_{ext}$ \citep{milgromQuasiN}.
Using $a_0 = 1.2\;\mathrm{\AA}\;\mathrm{s}^{-2}$ \citep{BBS}
and $g_{ext} = V_{\mathrm{MW}}^2/D$ with 
$V_{\mathrm{MW}} = 210\;\mathrm{km}\,\mathrm{s}^{-1}$ \citep{XueMW,mcgMW} at galactocentric 
$D = 19\;\mathrm{kpc}$ \citep{bellazzini}, the mass-to-light ratio of Sagittarius is 
$\Upsilon_*^V \approx 1.2\;\mathrm{M}_{\sun}/\mathrm{L}_{\sun}$ for the total luminosity estimated
by \citet{ibata}.  This is surprisingly reasonable,
but would be reduced if we adopted the larger luminosity estimate of \citet{NewSag}.
Perhaps this reasonable seeming mass-to-light ratio
is a fluke since all the same assumptions apply to the analysis here as in the Newtonian
case: stability and sphericity.  Neither condition is satisfied in Sagittarius, which is highly 
elliptical and far from equilibrium.

For the classical dwarfs excluding Draco and Ursa Minor, the inferred mass-to-light ratios are 
reasonable for stellar populations \citep{angusdw,serradw}.  
\citet{hernandezmond} even find that lower mass-to-light ratios tend to occur in dwarfs with
more recent star forming events, as expected from a stellar populations perspective.
This is not trivial, as the mass-to-light ratio computed in MOND is \textit{extremely}
sensitive to the accuracy of the data simply because the mass depends on a 
high power of the velocity dispersion.  Even small errors in identifying members 
can have a noticeable effect \citep{serradw}.  So far, however, Draco and Ursa Minor 
persist\footnote{The problem in these cases is lessened but not eliminated if we adopt the lower
velocity dispersions of \citet{walker}.} in being problem cases with uncomfortably high
($\sim 10\;\mathrm{M}_{\sun}/\mathrm{L}_{\sun}$) mass-to-light ratios \citep{GS92,milg7dw}.

For the ultrafaint dwarfs, the mass-to-light ratios are not at all reasonable \citep{SSH}.
Though the BTFR is no longer absolute when the external field dominates, it should remain
a reasonable proxy for MOND \citep{milgromQuasiN}.  In this approximation, the 
inverse of the deviations $F_b^{-1}$ are proxy estimates for the mass-to-light ratios in 
MOND\footnote{In this regard, we confirm the results of \citet{SSH} for the MOND 
mass-to-light ratios of these dwarfs.} as they measure how remote each dwarf is from the 
mean $\Upsilon_*^V = 1.3\;\mathrm{M}_{\sun}/\mathrm{L}_{\sun}$ of \citet{mario} 
and \citet{martinmaxL}.  From this perspective, $\Upsilon_*^V \approx {F_b}^{-1} > 10$ is 
quite unreasonable.  The cases with ${F_b}^{-1} \approx 100$ are simply absurd.  

The ultrafaint dwarfs may therefore be fatal to MOND, \textit{provided} that both their data
and the assumptions underlying the analysis are valid (\S~\ref{accuracy}).  
The assumption of sphericity is obviously not satisfied in many of the most deviant cases, 
though it seems unlikely that geometric corrections alone could
explain the large deviations seen in these objects.  The assumption that the dwarfs are 
in dynamical equilibrium is perhaps more consequential.  If they are, they should not exhibit 
discrepancies of one or two orders of magnitude.  If they are not, then MOND may become
similar to the stellar stripping hypothesis (\S~\ref{starstrip}), though not identical.
Tidal forces are stronger in MOND than in conventional gravity \citep{bradawarp} as the long 
range boost to the effective force becomes more important than the extra dark mass \citep{angmcg}.
This opens the possibility that stripping could be occurring at present.

We can estimate the tidal radii of the dwarfs in MOND just as we did conventionally in
\S \ref{starstrip}.  \citet{zhaotian} find that the expression for the tidal radius in MOND is nearly 
identical to that in the conventional case (equation~\ref{RTconventional}), differing only by
a factor of $(2/3)^{1/3}$:
\begin{equation}
r_{t,M}  = D \left(\frac{m}{2M}\right)^{1/3}
\label{RTMOND}
\end{equation}
What is really different here are the masses $m$ and $M$ of the satellite and host.
Here we simply set $m = M_b$, the baryonic mass of each dwarf estimated from its luminosity
as before. For the host mass $M$, we take
$M_{\mathrm{MW}} = 6 \times 10^{10}\;\mathrm{M}_{\sun}$ for the Milky Way \citep{flynn,mcgMW}, and
$M_{\mathrm{M31}} = 2 M_{\mathrm{MW}}$ for M31 \citep{M31RC}.  Note that these are the baryonic 
masses of the two host galaxies, as there is no dynamical dark matter in MOND.

Contrary to the case for dark matter, the residuals from the BTFR correlate strongly with the 
size of the dwarfs as measured by MONDian tidal radii (Fig.~\ref{tidalR}).
The closer the half light radius is to the tidal radius, the further the dwarf deviates from the BTFR.
Intriguingly, the size scale at which the deviation occurs corresponds well to the photometrically
measured tidal radius:  $r_{t,M} \approx r_{t,phot}$.  Galaxies that adhere to the BTFR are safely
within their tidal radii such that $r_{t,phot} < r_{t,M}$ while the deviant cases typically have
$r_{t,phot} \gtrsim r_{t,M}$.  Dwarfs with a larger fraction of their stars exceeding the tidal radius appear to
have lost more light.  This 
implies that tidal disruption may be the cause of the deviance of Draco, Ursa Minor, and 
the MW ultrafaint dwarfs.  The MONDian interpretation of these systems would thus appear to be
that they are in the process of being tidally distorted and ultimately shredded in the field of the 
Milky Way.  The small $F_b$ are not large MONDian $M_*/L_V$ because the assumption of
stable equilibrium does not hold.  Rather, a small ${F_b}$ is an indicator of the degree of tidal disruption.

The deviations $F_b$ in Fig.~\ref{tidalR} appear to follow a straight line that continues from well below
the BTFR to a bit above it.  Satellites should not have $F_b > 1$ as a result of tides in MOND.
Rather, once objects become isolated enough, they should adhere to the BTFR with no net residuals.
So this plot should go up to unity, then remain there.
Presumably this is what we would see if we could include the isolated disk 
galaxies\footnote{We do not, at present, have sufficient information to compute $g_{ext}$ for the
spirals in Fig~\ref{MbV}.} from Fig.~\ref{MbV}.  The perception of a linear rise to $F_b > 1$ is due to
three satellites of M31: And II, III, and VII.  For these three objects, the implied MOND mass-to-light
ratios are uncomfortably low: $\Upsilon_* \approx 0.2\;\textrm{M}_{\sun}/\textrm{L}_{\sun}$ would be
required to bring them onto the BTFR.  Given the uncertainties in both the analysis and the data 
(\S~\ref{accuracy}), we hesitate to put too much weight on these few cases.

The correlation seen in the MONDian portion of Fig.~\ref{tidalR} is quite strong.
With the exception of the most extreme outlier, it is consistent with little intrinsic scatter.
How great the intrinsic scatter is depends sensitively on what we take for the uncertainty 
in $\Upsilon_*$.  We would not expect a perfect correlation, as there
should be some variation due to the orbits of the dwarfs.  The expression for the tidal radius
implicitly assumes circular orbits; more generally we should replace 
$D \rightarrow \mathrm{a}(1-\mathrm{e})$.  Dwarfs on radial orbits should be more
susceptible to disruption, while retrograde orbits may be somewhat more stable.
While it is probably an over-interpretation of the data, we should at least point out that
this effect might be perceptible.  Two or three of the ultrafaint dwarfs (Canes Venatici II, Willman 1, 
and perhaps also the most discrepant case, Segue 1) sit below the main correlation.
One might infer from this that they are on more eccentric orbits than the other dwarfs.
Sextans, on the other hand, remains consistent with the BTFR in spite of having a half light
radius approaching its tidal radius.  This situation might persist longer if it happens to be on a 
retrograde orbit.  Another possibility is that Sextans is currently making its first close approach, 
so has not yet suffered disruption.  Indeed, \citet{bradadwarf} show that dwarfs on mildly elliptical
orbits may not disrupt at all if they are only exposed to pronounced tidal effects for a brief period 
near the pericenters of their orbits.

The numerical simulations of \citet{bradadwarf} show that disrupting satellites form stellar streams
(see their Fig.~5).  Thus the MOND hypothesis for explaining deviations 
from the BTFR becomes very similar to the stellar stripping hypothesis (\S~\ref{starstrip}). 
An important difference is that we need not invoke highly eccentric orbits for the dwarfs.  
Since $r_{t,M} \approx r_{t,phot}$, stellar stripping may be ongoing at present.  

The severity of the effect of the external field can be quantified \citep{bradadwarf} by 
\begin{equation}
\gamma = \left(\frac{D}{r}\right)^{3/2} \left(\frac{m}{M}\right)^{1/2}.
\label{mondadiabat}
\end{equation}
This measures how adiabatic the effects of the external field are.  
In effect, $\gamma$ is the number of orbits a star should make within a dwarf for every orbit the dwarf
makes about its host.  For specificity, we compute $\gamma$ for a star at
the deprojected 3D half light radius $r_{1/2}$ with $m = M_b/2$.
Equation~(\ref{mondadiabat}) applies when the field of the host dominates: 
$m/r^2 < M/D^2$.  Based on their baryonic masses, all dwarfs are in this 
regime\footnote{If instead we use $\sigma_{los}^2/r_{1/2}$ to estimate the internal acceleration,
all dwarfs are near the boundary $g_{int} \approx g_{ext}$ between domination by the internal and
external field.  However, we expect the velocity dispersion
to be inflated by the external field effect, so the apparent baryonic mass is the better indicator of regime
\citep[see][]{bradadwarf}.}
except Leo T.  For this galaxy, $\gamma = (D/r) (m/M)^{1/4}$ \citep{bradadwarf}.

In order for a dwarf to remain in equilibrium, there must be enough internal orbits for the
dwarf to adjust to variations in the external potential.  This raises the question of how many
orbits suffice to maintain an adiabatic variation.  Certainly there should be at least two
internal orbits per significant change in the potential.  The potential of the host is due
entirely to the baryonic mass as there is no dark matter halo in MOND, so a 
dwarf  on a polar circular orbit sees the host vary from face-on to edge-on in one
quarter of an orbit.  This is a significant variation in the potential that will only be magnified for
non-circular orbits.  Combining these criteria, we suggest that a minimum value 
for the adiabatic condition to hold is $\gamma > 8$.  This crude argument appears to be consistent with the
numerical simulations of \citet[see their Fig.~4]{bradadwarf}:  satellites that remain distant enough
from their host to always have $\gamma > 8$ experience only small perturbations.
Satellites reaching $\gamma \lesssim 8$ near the pericenters of their orbits 
puff up in size and decline in velocity dispersion, but recover as they recede. 
Satellites that enter the regime $\gamma < 2$ experience unchecked growth in size,
subjecting them to tidal disruption and the formation of tidal streams.  The velocity dispersions 
of these systems initially decline but then grow.  This is qualitatively the right vector
for an evolutionary track that would explain the deviations from the BTFR, 
but considerably more work would be required to predict such a track or even to check 
how sensitive the evolution is to parameters such as orbital eccentricity.

\begin{figure*}
\epsscale{1.0}
\plotone{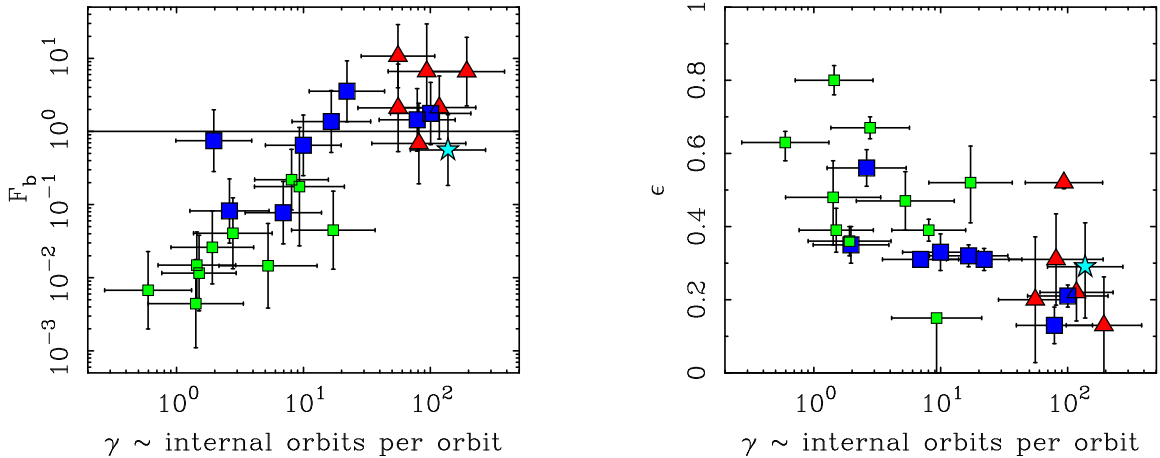}
\caption{The residuals from the BTFR (left) and the ellipticities of dwarfs (right;
symbols as per Fig.~\ref{MbV})
correlate with the number of orbits a star should complete within a dwarf for every orbit the 
dwarf completes about its host (equation \ref{mondadiabat}).
We expect pronounced non-adiabatic effects when $\gamma \lesssim 8$ (see text).
This corresponds approximately to where the discrepancy from the BTFR
becomes significant, and to where dwarfs tend to become non-spherical.
\label{orbits}}
\end{figure*}

Fig.~\ref{orbits} shows the deviations from the BTFR and the ellipticity of the dwarfs
plotted as a function of $\gamma$ computed at the half light radius.  Correlations with both
are apparent.  The deviation sets in around $\gamma \approx 10$, consistent with the
above reasoning.  The most distorted dwarfs are those with the least time to react
to changes in the external field.  Dwarfs with large $\gamma$ have many orbits to adjust to 
changes in the potential; these objects are mostly round.  While it may be possible to avoid
distorting the shapes of dwarfs via tides when they are protected by a cocoon of dark matter
\citep{munozcar}, tidal distortion seems inevitable in MOND \citep{bradadwarf}.

\placefigure{orbits}

We note that the plot of the deviations in Fig.~\ref{orbits} is very similar to Fig.~\ref{tidalR}.  
One interesting difference is the case of Leo T, which is in the pure MOND regime. 
Consequently, it executes fewer internal orbits per external orbit than it would if it
were in the quasi-Newtonian regime like the rest of the dwarfs.  This moves it closer to the
clump of dwarfs that adhere to the BTFR in Fig.~\ref{orbits} than in Fig.~\ref{tidalR}. 
We also note that Leo T is the only dwarf considered here that retains gas.  Tides can
strip gas, with the consequences discussed in \S \ref{rampress}.  It may therefore be 
significant that all of the remaining dwarfs are dominated by the external field of the host.
The absence of gas in dwarfs within $\sim 250$ kpc might be a signature of the transition
to domination by the external field in MOND.

The structure of Figs.~\ref{tidalR} and Fig.~\ref{orbits}
should be similar in MOND, but this behavior is not guaranteed.  It only follows if the
MOND formula for $\gamma$ (equation~\ref{mondadiabat}) provides the correct timescale.
This would not follow had we used a Newtonian estimator for the timescale, either with dark matter or without.
With dark matter, orbits within the dwarfs are more rapid thanks to the extra mass, and they always
have time to equilibrate ($\gamma \gg 10$).  Without dark matter in the dwarfs themselves, the dark
matter halos of the hosts dominate and the dwarfs should dissolve rapidly ($\gamma \ll 1$).
Only when we use the baryonic mass for both dwarf and host in equation~(\ref{mondadiabat})
does the onset of the deviation from the BTFR correspond to when internal and external orbital
timescales are comparable.
This is one sense in which the external field effect is unique to MOND,
and appears to be consistent with the data for the MW dwarf spheroidals.  

A separate question is whether the number of dwarfs currently near total dissolution is reasonable.
By reasonable, we mean a dissolution timescale that is shorter than a Hubble time so that there is time for
deviation from the BTFR to have occurred,  but not so short that we must be catching them at a special
time.  Unfortunately, it is not clear what the dissolution timescale is \citep[see][]{mondUMi}.
The trends in Fig.~5 are suggestive of progressive mass loss, but we do not know what the mass loss
rate is, so cannot quantify $t_{dissolve} = |M_b/\dot M_b|$.  As a proxy, we can estimate the crossing time 
$t_{cross} = 2r_{t,phot}/\sigma_{los}$.  Presumably it takes a number of crossing times for a dwarf to
dissolve, though it is unclear how many.  Most dwarfs have $10^8 < t_{cross} < 10^9$ years.  
The MW satellite with the longest crossing time is Sextans ($\sim 1$ Gyr), which is a fair fraction
of its orbital period ($\sim 3$ Gyr if $\textrm{a} = D$).  This object provides an interesting constraint,
as it has had time to complete four circular orbits and a dozen crossing times in a Hubble time.
This seems like plenty of time for a low $\gamma$ system so close to its tidal radius to have suffered
some damage, and yet it remains consistent with the BTFR.  If instead it is in a somewhat eccentric
orbit that only now places it in the regime of tidal susceptibility, its crossing time is long enough that
perhaps we do not (yet) see deviation in this case simply because the system has not had
time to react.  On the opposite extreme, the smallest of the ultrafaint dwarfs, Segue 1 and Willman 1, 
have the shortest crossing times: $t_{cross} \approx 5 \times 10^7$ years.  This is uncomfortably short.
Both are far from the BTFR, so perhaps have been caught just before total dissolution.  Nevertheless,
these cases would appear to pose a dissolution timescale problem.  It is hard to say how severe this
problem is without knowledge of the initial population.  The mass function is not known in MOND,
nor is it even clear whether the dwarfs are primordial or tidal in origin \citep{gentiletidal}.

The properties of the deviant dwarfs provide a strong test of the MOND hypothesis.
If they are stable, undisturbed systems, then their mass-to-light ratios are unacceptably large and
MOND fails.  If instead they are being tidally stripped, this situation is naturally understood in the context of
MOND for which the tidal radii are comparable to the luminous size.  This differs from the stellar stripping
hypothesis in the dark matter context in that tidal disruption should be going on now for the deviant dwarfs,
and not just when the dwarfs are near the pericenters of their orbits.  In addition, the fact that the tidal radii 
of the dwarfs seem to be adequately estimated by assuming $\mathrm{a} \approx D$ implies that their orbits 
are not highly eccentric.  The distribution of orbital eccentricities of the dwarfs may therefore provide another
test to distinguish MOND from \LCDM.

\section{Conclusions}

We have examined the adherence of the Local Group dwarf satellites of the Milky Way and M31
to the Baryonic Tully-Fisher Relation.  We find that most of the brighter dwarfs are largely consistent with the
extrapolation of the BTFR fitted to isolated, late type, gas rich, rotating disk galaxies.  
The fainter dwarfs, especially the ultrafaint dwarfs, are not.  More importantly, we find that
residuals from the BTFR are not random, correlating well with luminosity and ellipticity.  
The amount of deviation from the BTFR also correlates with metallicity, size, 
and the susceptibility of the dwarfs to tidal perturbation.
We have considered a number of possible interpretations for the observed behavior, 
as we summarize below.

\paragraph{Insufficient Kinematic Accuracy:}
Heroic efforts have been made to find new dwarfs and to measure their velocity dispersions.
This is a challenging endeavor.  While the deviations of some dwarfs from
the BTFR are formally significant, that significance is not overwhelming (typically 2 to $4 \sigma$).
Moreover, the analysis assumes that the dwarfs are spherical and in stable equilibrium.
The assumption of sphericity at least is violated for the most deviant dwarfs.  We therefore consider
one possibility to be that there are no genuine deviations from the BTFR.  This hypothesis predicts
that as the data improve, so too will agreement with the BTFR.

\paragraph{Gas Removal:}
Dwarfs that deviate from the BTFR do so in the sense that they seem to be lacking luminosity 
for their velocity dispersion.  This may be explained if baryons are removed before they form stars.  
Several possible mechanisms to accomplish this removal include the suppression of star formation 
by cosmic reionization, ionization from Pop.~III stars, removal of cold gas by
supernova feedback, and ram pressure stripping.  

Cosmic reionization is often invoked in the context of the dwarf spheroidals, 
and is an attractive solution if only these objects are considered.
If we simultaneously consider slightly larger gas dominated disk galaxies, it becomes clear 
that reionization is not in itself an adequate explanation for the observed trends in the data.  
Some other mechanism must be acting to suppress the accumulation of cold baryons in a manner
that becomes more severe with decreasing $V_c$.  Whatever this mechanism is, it presumably
affects the dwarf spheroidals as well.  Cosmic reionization may be an additional factor acting
only at scales $V_c < 20\;\mathrm{km}\,\mathrm{s}^{-1}$.

Feedback from supernovae is a candidate mechanism for affecting star formation across all halo
masses. This provides a qualitatively appealing explanation for the trend in the detected baryon 
fraction with halo mass.  Supernovae provide the kinetic energy to drive gas beyond
escape velocity, but the escape velocity increases with increasing halo mass so progressively
more baryons are retained.  This mechanism may provide a natural explanation for the residual
correlation with luminosity and metallicity.  However, quantitative tests remain wanting, as
does an explanation for the correlations of the residuals with ellipticity and tidal susceptibility. 

In the ram pressure stripping hypothesis, galaxies deviate from the BTFR when they fall into
the halo of the current host galaxy and their cold gas is ablated by the ram pressure of the 
hot gas in the halo.  This hypothesis requires that a sufficient amount of hot gas be present
in the halos of the host galaxies, which is not obviously the case.  
It predicts that the ages of stars in the dwarfs is related to the time
of infall, with the dwarfs that fall in first losing the most gas, deviating by the largest amount from the
BTFR, and having the oldest stars.  Strictly speaking, the star formation history of the pre-infall
dwarf is not constrained, so there should be some scatter in these predictions.  However, no
star formation can occur after infall and gas stripping, so the ages of the youngest stars should
correspond well to the time of infall with a sharp truncation in star formation after that time.  
This would provide a natural explanation for why some dwarfs appear to have had relatively
recent star forming events but now contain no cold gas.  One would also expect the metallicities of 
the stars to reflect the star formation history.  The first dwarfs to fall in would have had the least time
for enrichment and have the lowest [Fe/H].  This predicts that [$\alpha$/Fe] should also correlate
with the amplitude of deviation $F_b$, to the extent to which we expect the objects with the briefest 
enrichment time to have the highest [$\alpha$/Fe].

The gas removal hypotheses provide a potential explanation for the correlation of BTFR residuals
with luminosity and metallicity.  However, they provide no obvious explanation for the correlation
with ellipticity and tidal susceptibility.  This occurs more naturally in the following two scenarios.
The removal of gas and the concomitant truncation of star formation and its consequences for metal 
enrichment may also occur as a result of tidal stripping in the following hypotheses.

\paragraph{Stellar Stripping:}
The correlation of the residuals with ellipticity in addition to luminosity suggests a role for tidal effects.
In this hypothesis, the dwarfs become progressively more distorted due to tidal disruption as they 
orbit the Milky Way.  Stars are lost in the process, reducing the luminosity of the dwarfs.  
At the present time, the computed tidal radii of the dwarfs greatly exceed their luminous extent.
This leads us to infer that the orbits of the dwarfs must be highly eccentric in this scenario, with most
of the stripping occurring during pericenter passage.  This hypothesis predicts that the mass required to 
reconcile each dwarf with the BTFR may exist in a tidal stream.  It further predicts that the age and
metallicities of stars in the predicted streams should be consistent with those of the parent body.
Examples exist where this may already be observed.

\paragraph{MOND:}
Low surface brightness dwarf spheroidals provide a strong test of an alternative to dark matter, MOND.
They should very nearly follow the BTFR, which is a consequence of the specific form of the
modified force law in MOND.  While some dwarfs do indeed adhere to the BTFR, others 
deviate substantially.  The ultrafaint dwarfs of the Milky Way have
MONDian mass-to-light ratios in the tens to hundreds.  This is fatal for MOND \textit{if}
these dwarfs are in a stable equilibrium, the spherical approximation used in the analysis is
adequate, and the kinematic data are to be trusted.  Intriguingly, the sizes of the dwarfs relative to their
MONDian tidal radii correlate strongly with the degree of deviation from the BTFR.  Indeed,
the discrepancy for MOND sets in precisely where the theory predicts that non-equilibrium
effects become strong.  It therefore 
appears that the unacceptably high mass-to-light ratios may be a result of the dwarfs being
out of equilibrium.  This should be testable, in the sense that the deviant dwarfs should show
evidence of tidal disruption while the dwarfs that adhere to the BTFR should not.  
Notably, stripping of the deviant dwarfs should be ongoing and not restricted to pericenter passage
as in the stellar stripping hypothesis.

It is of course possible that some combination of these effects is at work.  As a dwarf satellite
approaches its host on its orbit, it is subject to both tidal forces and ram pressure effects.
Perhaps gas is lost first due one or both of these effects, with stars being tidally liberated later.  
This makes it somewhat difficult to distinguish between the various hypotheses, but it should
be possible.  

\acknowledgements  The work of S.S.M.\ is supported in part by NSF grant AST 0908370.  
We thank Oleg Gnedin for organizing the workshop where this work was conceived, and the
referee for a thorough reading and many insightful comments.
We also thank Matt Walker, Michele Bellazzini, Xavier Hernandez, Rosemary Wyse, Mario Mateo, 
Mia Bovill, Chris Mihos, Moti Milgrom, Massimo Ricotti, and Cole Miller for related conversations.

\end{document}